%% file: ms.tex
\documentclass{emulateapj}

\begin{document}

\newcommand{\kms}{\,km\,s$^{-1}$}
\newcommand{\vej}{v_{\mathrm{ej}}}
\newcommand{\vlsr}{v_{\mathrm{LSR}}}
\newcommand{\acd}{N_{\mathrm{a}}}
\newcommand{\zabs}{z_{\mathrm{abs}}}
\newcommand{\zem}{z_{\mathrm{em}}}
\newcommand{\aox}{\alpha_{\mathrm{ox}}}
\newcommand{\cloudy}{{\sc cloudy}}


\title{High-Resolution Absorption Spectroscopy of Multi-phase, High-Metallicity
Gas Associated with the Luminous Quasar HE\,0226-4110
\footnote{Based on observations made with the NASA/ESA Hubble Space
Telescope, which is operated by the Association of Universities for
Research in Astronomy, Inc., under NASA contract NAS 5-26555. Also
based on observations made with the NASA-CNES-CSA Far Ultraviolet
Spectroscopic Explorer, which is operated for NASA by Johns Hopkins
University under NASA contract NAS 5-32985.}}

\author{Rajib Ganguly\altaffilmark{1,2},
        Kenneth R.~Sembach\altaffilmark{1},
        Todd M.~Tripp\altaffilmark{3},
        Blair D.~Savage\altaffilmark{4},
        Bart P.~Wakker\altaffilmark{4}}
\altaffiltext{1}{Space Telescope Science Institute, 3700 San Martin
Drive, Baltimore, MD  21218}
\altaffiltext{2}{Department of Physics \& Astronomy, University of
Wyoming (Dept. 3905), 1000 East University Avenue, Laramie, WY
82071}
\altaffiltext{3}{Department of Astronomy, University of
Massachusetts, Amherst, MA 01003}
\altaffiltext{4}{Department of Astronomy, University of
Wisconsin-Madison, 475 N. Charter Street, Madison, WI  53706}

\begin{abstract}
We present FUSE and HST/STIS observations of the absorption line
system near the emission redshift of the radio-quiet, X-ray bright
quasar HE0226-4110 ($z = 0.495$, $V=15.2$). The spectra cover the
rest-frame wavelength range 610--1150\,\AA, and we detect a wide
range of ionization species, including four adjacent stages of
oxygen: {\ion{O}{3}}, {\ion{O}{4}}, \ion{O}{5}, and \ion{O}{6}.
Strong transitions of {\ion{O}{1}} and \ion{O}{2}\ are covered in
our spectra, but none are detected. The detection of multiple
ionization stages of a common element (oxygen) reveals a striking
change in gas kinematics with ionization. Comparison of the
{\ion{O}{6}}$\lambda\lambda$1031, 1037 apparent column density
profiles reveals no evidence for partial coverage or unresolved
saturated structure, although parts of the \ion{O}{6}$\lambda$1037
line are blended with Galactic \ion{C}{4} absorption. In addition,
several transitions (e.g, \ion{C}{3} $\lambda$977, \ion{H}{1}
Ly-$\beta$) show black saturation which also indicates no unocculted
flux that may dilute the absorption profiles. \ion{O}{3}\ is only
detected in a narrow feature which is also traced by the \ion{H}{1}\
and \ion{C}{3}\ lines, suggesting that they arise in the same gas.
Absorption at the same velocity is also present in other species
(\ion{N}{4}, \ion{O}{4-VI}, \ion{S}{4}, and possibly {\ion{Ne}{8}}),
but the kinematics differ from the \ion{O}{3}, implying production
in separate gaseous phases. The combination of \ion{H}{1},
\ion{O}{3}, and \ion{C}{3} column density measurements with limits
on the amount of \ion{O}{2}\ and \ion{O}{4} in this phase yield an
estimate of both the photoionization parameter and metallicity:
[O/H]$=+0.12_{-0.03}^{+0.16}$, $\log U=-2.29_{-0.23}^{+0.02}$. We
discuss two possible locations for the gas in this associated
absorption-line system: the narrow emission line region of the
quasar, and the halo of the quasar host galaxy. An additional narrow
component that is only detected in \ion{O}{6} appears 58\,\kms\
redward of the \ion{O}{3}-bearing gas. The narrow width of this
component rules out collision-based ionization processes, and we
constrain the ionization parameter of this component to $-0.35
\lesssim \log U \lesssim 0.02$. Additional structure is detected in
the associated absorber in the form of two broad components. The
components flank the narrow \ion{O}{3} component in velocity and are
detected in \ion{N}{4}, \ion{O}{4-VI}, and \ion{Ne}{8}. The
kinematics of {\ion{O}{5}} and \ion{O}{6}\ in both components trace
each other. The kinematics of \ion{N}{4}, \ion{O}{4}, and
\ion{Ne}{8} differ from \ion{O}{5}\ and \ion{O}{6} and it is not
clear how many phases or what ionization mechanism produces the gas
in these broad components. A consideration of collisional ionization
and photoionization equilibrium models, as well as radiative cooling
and shock ionization models, leads us to conclude that \ion{Ne}{8}
must arise in a separate high-ionization phase.
\end{abstract}


\section{Introduction}

Early after the discovery of quasars, it was recognized that their
extreme luminosities, non-thermal spectra, variability, and compact
sizes were readily achieved through the release of gravitational
energy of matter accreting onto supermassive black holes
{\citep[e.g.,][]{gs64,schmidt63}}. The remarkable result from the
previous decade of work that nearly all galaxies host central
supermassive black holes
{\citep[e.g.,][]{tremaine02,nuker00,fm00,magorrian98}} seems to
suggest that any galaxy has the potential to host a quasar (or other
form of active nucleus). In tandem with the observed fact that the
quasar luminosity function peaks at $z \sim 2$
{\citep[e.g.,][]{boyle00}} and falls off thereafter suggests that
the quasar phenomenon is possibly a phase of galaxy evolution.
Consequently, in order to trace the history of gas in the universe,
it is important to understand how accretion disks around
supermassive black holes are fueled, the dynamics and geometry of
the accreting/outflowing gas, and how some fraction of the gas is
returned to the IGM (i.e. feedback). A complete description of how
gas falls into galaxies and is subsequently processed should include
an understanding of the physical origin of absorption line systems
that appear to cluster around quasars \citep[e.g., ][hereafter
``associated'' absorption lines, or AALs\footnote{Following
\citet{foltz86}, we refer to any narrow absorption line that arises
within 5,000\,\kms\ as an AAL. We distinguish this purely
observational classification from absorption that is truly related
(i.e., intrinsic) to the quasar central engine or host
galaxy.}]{wwpt,foltz86,foltz88}. Are there dynamical/kinematic
relationships between the gas and those regions? Where is the gas
located? What are the physical properties of the gas? Are AALs just
weather around quasars or do they probe physically interesting
regions around the black hole like the broad/narrow emission line
region? Are AALs an integral feature of the quasar phenomenon?

The current paradigm of quasars holds that the disk of matter
accreting onto the central supermassive black hole that powers
quasars is also the origin of a wind
{\citep*[e.g.,][]{elvis00,psk00,mur95}}. This wind is thought to
result from gas that is lifted off the accretion disk via local
radiation pressure and then radially driven away from the black hole
via scattering of ultraviolet lines. This scenario has been quite
successful at explaining the existence and shape of both broad
absorption lines (BALs) and the high-ionization broad emission
lines. {Low-ionization and high-excitation lines are more
complicated as these may have a contribution from the outer regions
of the accretion-disk itself {\citep[e.g.,][]{eh04}}. There are
indications, including temporal variability of profiles and partial
occultation of the absorber over the background quasar, that some
fraction of narrow absorption-line systems also probe this
outflowing gas \citep[e.g., ][]{gan03b,misawa03,yuan02}.

In sight lines to high-redshift QSOs ($z \sim 2.5$) that are
selected based on radio brightness, as many as 30\% of
\ion{C}{4}-selected systems with $\zabs \ll \zem$, commonly assumed
to arise from the halos of intervening galaxies, may be intrinsic to
the QSO \citep{rich99,rich01b}. This frequency was estimated by
examining differences in the distribution of \ion{C}{4}-selected
systems in apparent ejection velocity (over the range $+75,000 \geq
\vej \geq -6,000$\,\kms) as a function of quasar radio loudness and
spectral index. If \ion{C}{4}-selected systems are unrelated to the
background quasar (as in the case of halos from
cosmologically-distributed interloping galaxies), then no variation
is expected. In the apparent ejection velocity range +6,000 to
+75,000\,\kms\ (i.e., at large separations from the quasar
redshift), \citet{rich01b} found a statistical excess of
\ion{C}{4}-systems toward radio-quiet quasars compared to radio-loud
quasars, and in flat-spectrum quasars compared to steep-spectrum
quasars.

This estimate of the frequency of high-ejection velocity systems is
surprisingly high given that, at lower redshifts, many studies have
shown that absorption systems with $\zabs \ll \zem$ are strongly
correlated with foreground galaxies, indicating that these low$-z$
absorbers are {\it not} intrinsic systems with high ejection
velocities. It is not yet clear how these observations can be
reconciled. Studies to understand the populations of absorbing
galaxies at high-redshift (i.e, $\zabs \gtrsim 1.5$) are not yet
feasible with current technology. Likewise, it is possible that the
selection of optically-bright low-redshift quasars in follow-up
studies of absorbing galaxy populations has an effect on the
velocity distribution of intrinsic systems such that they are not
observed.

In a heterogeneous study of low-redshift quasars, \citet{wise04}
have found that at least 20\% of low$-z$ AALs (i.e, systems
appearing within 5000\,\kms\ of the quasar redshift) show time
variability and therefore are likely to be intrinsic. While
intrinsic absorbers that lie at large velocity separations from the
quasar likely originate in some form of high-velocity outflow, there
are additional possible locations for AALs: ablation off an
obscuring torus {\citep{kk01}}, the narrow emission line region, the
quasar host galaxy {\citep[e.g.,][]{baker02}}, satellite dwarfs, or
luminous companion galaxies in the quasar's host cluster.

There has been considerable statistical work on the relationship
between associated absorbers and various quasar properties.
\citet{foltz86,foltz88}\ established that intermediate redshift AALs
($1.43 \lesssim z \lesssim 1.94$) with large \ion{C}{4} equivalent
widths [W$_{\mathrm{rest}}$(\ion{C}{4})$\gtrsim 1.5$\,\AA]
preferentially appear toward steep-spectrum radio-loud quasars, in
stark contrast with BALs, which prefer more radio-quiet quasars
\citep[e.g.,][]{becker00}. While the dichotomy in radio-loudness in
quasars has apparently been eliminated through deep radio imaging by
the FIRST survey \citep{first}, there remains an unexplained
discrepancy between quasar radio-loudness and the presence of
intrinsic absorption. \citet{gan01a} reported an apparent dearth of
strong AALs among the lower redshift ($z<1$) quasars observed with
the {\it Hubble Space Telescope} (HST) for the HST Quasar Absorption
Line Key Project, suggesting possible evolution in the frequency of
strong AALs. While the absence of strong AALs in that sample might
be explained partially as a bias against bluer quasars, there is
still evidence for evolution from larger and well-selected quasar
samples \citep{brandt,ves03}.

\begin{figure*}
\epsscale{0.9}
\rotatebox{-90}{\plotone{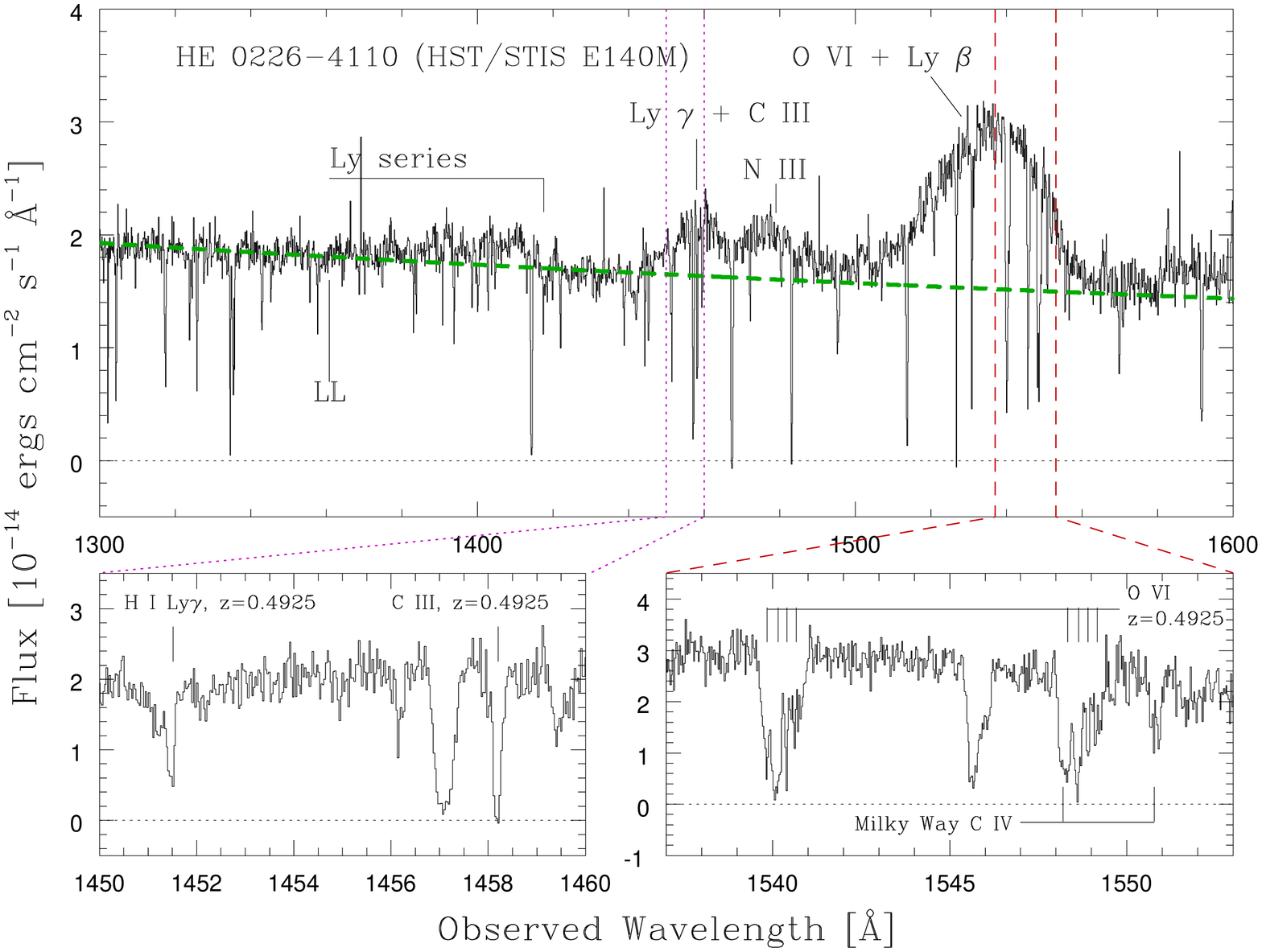}}
\caption{Portions of the STIS E140M echelle spectrum of HE0226-4110.
In order to improve the signal-to-noise ratio and thereby show the
continuum and emission-line shapes more clearly, the upper panel is
binned to a sampling of $\sim$20 km s$^{-1}$.  In the upper panel,
the thick dashed line shows the power law fitted to the continuum
for the purposes of the photoionization models presented in the text
(see \S \S 3.1,4.1). The \ion{O}{6}+Ly$\beta$, \ion{N}{3},
Ly$\gamma$+\ion{C}{3}, and higher order Lyman series emission lines
of the QSO are marked. An `LL' marks the location of the Lyman limit
for the associated absorption-line system. The lower panels show
selected associated absorption lines at $\zabs = 0.4925$ including
\ion{H}{1} Ly$\gamma$ and \ion{C}{3} $\lambda$977.020 (lower left
panel) and the \ion{O}{6} $\lambda \lambda$1031.926, 1037.617
doublet (lower right panel). The lower panels are binned to $\sim$7
km s$^{-1}$ pixels. Note that the \ion{O}{6} $\lambda$1037.617 line
is blended with one of the lines of the Galactic \ion{C}{4} $\lambda
\lambda$ 1548.204, 1550.781 doublet, which is also
marked.\label{fig:stisdata}}
\end{figure*}

As noted by several authors \citep[e.g.,][]{mew95,brandt}, there is
apparently a connection between \ion{C}{4}-selected AALs, and
absorption at soft X-ray wavelengths (so-called ``warm absorbers'').
The precise nature of this connection is perplexing given the
drastically different ionization potentials of species (e.g.,
\ion{O}{7}, \ion{O}{8}) that exist in warm absorbers. A number of
campaigns have been carried out employing simultaneous HST Space
Telescope Imaging Spectrograph (HST/STIS), {\it Far Ultraviolet
Spectroscopic Explorer} (FUSE), and {\it Chandra X-ray
Observatory}/XMM-Newton spectroscopy of Seyfert galaxies, the
lower-luminosity kin of quasars, such as NGC 3783
{\citep{ngc3783iv,ngc3783iii,ngc3783ii,ngc3783i}}, Mrk 279
{\citep{scott04}}, and NGC 5548 {\citep{cren03}}. The results of
these studies of Seyfert galaxies in regards to the general
relationship between the X-ray and UV absorbing gas is inconclusive.
In some cases, the absorbers can be explained by the same phase of
gas; in other cases, a more complex, multi-phase absorber is
required. For the more luminous quasar 3C\,351, \citet{yuan02} have
analyzed high resolution UV and X-ray spectra and find that complex,
multi-phase gas is required to explain the components in the AAL.

Like the previously mentioned campaigns, it is important to temper
statistical relationships between absorbers and quasar properties
with case studies of individual objects. Since the demographics of
the absorbers are clearly eclectic, only case-by-case
classifications to identify subsamples can hope to recover true,
undiluted physical relationships. Using HST/STIS and FUSE
observations of low-redshift quasars that cover the \ion{O}{6}
$\lambda\lambda$1031.927,1037.617 doublet, not only can we bridge
the gap between \ion{C}{4}-selected AALs and soft X-ray ``warm
absorbers,'' but we can also take advantage of the high spectral
resolution to understand the physical properties and origin of the
gas. \citet{kriss02}\ presents a summary of FUSE observations of
bright AGN at low-redshift (predominantly Seyfert galaxies). These
observations survey the \ion{O}{6} doublet in AGN out to $z \sim
0.15$. In a companion paper, we extend this survey out to $z \sim
0.5$ (Ganguly et al. 2005, in prep.).

In this paper, we report the discovery of a remarkable absorption
line system ($\zabs=0.4925$) near the redshift of the radio-quiet,
X-ray bright quasar HE\,0226-4110
\citep[$\zem=0.495$,$V=15.2$;][]{rei96}. In \S\ref{sec:data}, we
discuss the details of our HST/STIS and FUSE observations. In
\S\ref{sec:quasar}, we investigate the properties of the quasar to
provide a context for understanding the associated absorption
presented and analyzed in \S\ref{sec:aal}. Finally, in
\S\ref{sec:discussion}, we summarize our results and discuss the
implications of our findings.

\section{Data}
\label{sec:data}

\begin{figure*}
\epsscale{0.9}
\rotatebox{-90}{\plotone{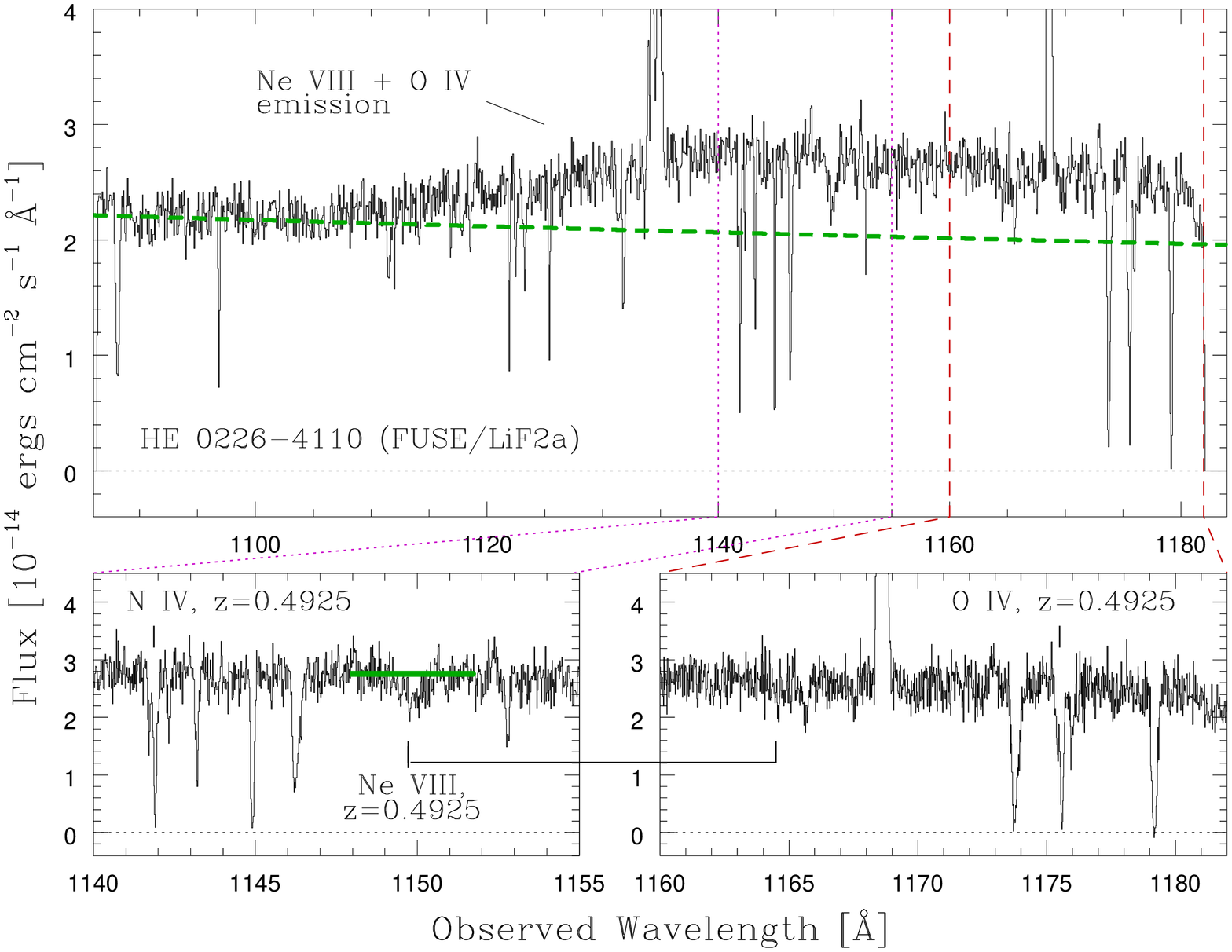}}
\caption{Portions of the FUSE LiF2A spectrum of HE0226-4110. In
order to improve the signal-to-noise ratio and thereby show the
continuum and emission-line shapes more clearly, the upper panel is
binned to a sampling of $\sim$20 km s$^{-1}$.  In the upper panel,
the thick dashed line shows the power law fitted to the continuum
for the purposes of the photoionization models presented in the text
(see \S \S 3.1,4.1), and the \ion{Ne}{8}+\ion{O}{4} emission line of
the QSO is marked.  The lower panels show selected associated
absorption lines at $\zabs = 0.4925$ including \ion{N}{4}
$\lambda$765.148 (lower left panel), \ion{O}{4} $\lambda$787.711
(lower right panel), and the \ion{Ne}{8} $\lambda\lambda$770.409,
780.324 doublet (both lower panels. In the lower left panel, our
local Legendre-polynomial fit to the continuum around the
\ion{Ne}{8} $\lambda$770.409 line is shown as a thick line. The
\ion{Ne}{8} $\lambda$780.324, which would nominally appear in the
lower right panel, is not detected. The lower panels are binned to a
sampling of $\sim$6 km s$^{-1}$. \label{fig:fusedata}}
\end{figure*}

\subsection{STIS Observations}

Our HST/STIS \citep{stis98,kimble98} observations were carried out
using the  $0.\!\!''2 \times 0.\!\!''06$\ slit, the E140M grating,
and the NUV MAMA detector between 25 December 2002 and 1 January
2003. The total integration time of the observations was 43.8\,ksec.
This configuration provides a spectral resolution of about 6.5\,\kms
(FWHM) with a sampling of 2--3 pixels per resolution element, and
yields semi-continuous spectra over the observed wavelength range
1149--1729.5\,\AA. The data were reduced and calibrated as described
by \citet{tripp01}. For further details on the processing of the
datasets used in this work for the HE\,0226-4110 sight line, see
\citet{fox05} and \citet{savage05a}. According to \citet{stis}, the
zero-point heliocentric velocity uncertainty in the calibration is
about 0.2--0.5 pixels, or {$\sim0.6-1.5$\,\kms}. The flux
calibrations are good to 8\% (or roughly $1-2 \times 10^{-15}$\,erg
cm$^{-2}$\ s$^{-1}$\ \AA$^{-1}$\ for these data). The integrations
yielded a final signal-to-noise per resolution element (after
reduction, calibration, and co-addition of spectra) of $\sim 11$\ at
1300\,\AA\ and $\sim 8.5$\ at 1500\,\AA.

Selected portions of the STIS echelle spectrum of HE\,0226-4110 are
shown in Figure~\ref{fig:stisdata}. The upper panel of
Figure~\ref{fig:stisdata} shows a broad section of the spectrum
including the \ion{O}{6}+Ly$\beta$ broad emission line of the QSO as
well as several weaker emission features. We also show in the upper
panel a power law fitted to the QSO continuum (see
\S\ref{sec:uvopt}) which we will use for photoionization modeling in
\S\ref{sec:narrow}. The spectrum is rich in Galactic and
extragalactic absorption lines. Extragalactic intervening absorption
systems (i.e., with $\zabs \ll \zem$) in the spectrum of
HE\,0226-4110 are analyzed and discussed in \citet{savage05a} and
\citet{lehner05}, and Galactic absorption lines and high-velocity
clouds toward this QSO are discussed by \citet{csg05a} and
\citet{fox05}.

\begin{figure*}
\epsscale{0.55}
\rotatebox{-90}{\plotone{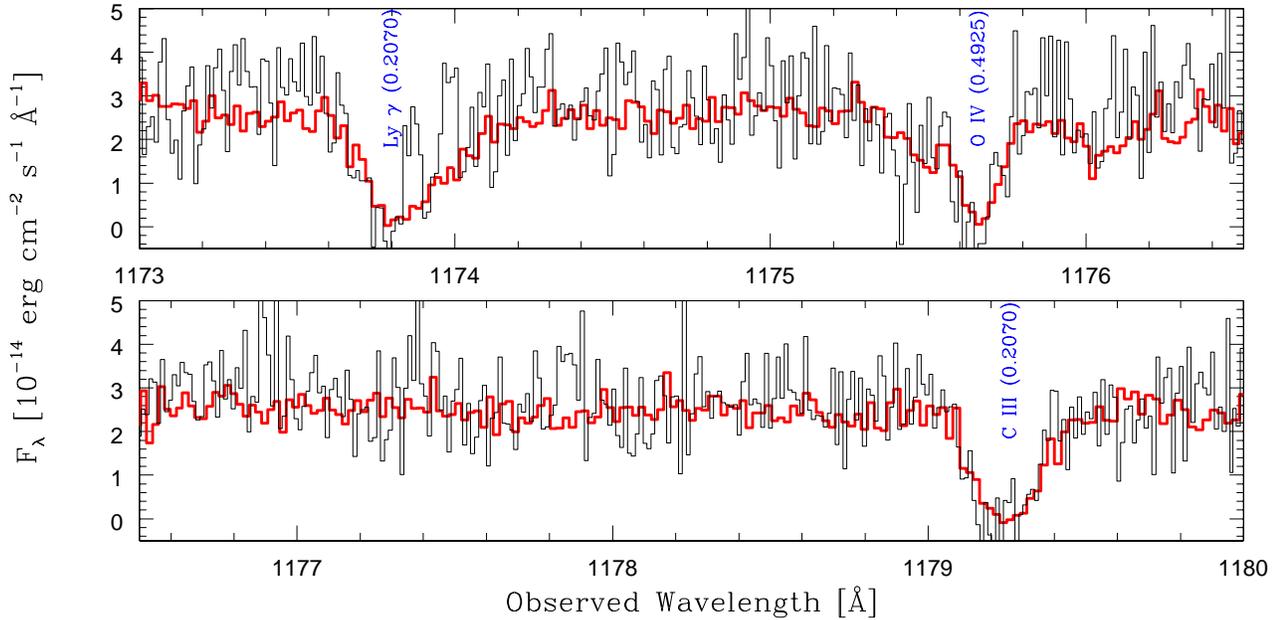}}
\protect\caption[FUSE/STIS Alignment]{We show a portion of the
overlapping region between the FUSE LiF2A (thick histogram) and STIS
E140M (thin histogram) spectra. The plotted region contains
associated absorption from {\ion{O}{4}}, as well as IGM absorption
from Lyman $\gamma$\ and {\ion{C}{3}}.} \label{fig:alignment}
\end{figure*}

Examples of the associated absorption lines of interest in this
paper that are redshifted into the STIS-E140M band are shown in the
lower panels of Figure~\ref{fig:stisdata} including the
multicomponent \ion{O}{6} $\lambda \lambda$ 1031.926, 1037.617
doublet at $\zabs$ = 0.4925 (lower right panel) and the \ion{H}{1}
Ly$\gamma$ and \ion{C}{3} $\lambda$977.020 lines (lower left panel).
It is immediately evident that this is a complex, multiphase
absorber: four components are present in the \ion{O}{6} lines spread
over a velocity range of 250\,\kms, but only a single component is
seen in tracers of lower-ionization gas such as \ion{C}{3} and
Ly$\gamma$ (the weak component blueward of the Ly$\gamma$ line shown
in Figure~\ref{fig:stisdata} is an unrelated Ly$\alpha$ line at a
different redshift). We also see from this figure that the
associated \ion{O}{6} $\lambda$1037.617 line at $\zabs$ = 0.4925 is
blended with Milky Way \ion{C}{4} $\lambda$1548.204.  This blend is
not a serious problem because the \ion{O}{6} $\lambda$1031.926 line
is free of blending, and the Galactic \ion{C}{4} $\lambda$1550.781
transition is also unblended, which enables an assessment of the
degree of contamination from \ion{C}{4} $\lambda$1548.204.

\subsection{FUSE Observations}

Our FUSE observations were carried out using the LWRS aperture
between 12 December 2000 and 21 October 2003 for a total integration
time of 208.9\,ksec. In this mode, the FUSE satellite delivers
spectra with velocity resolution in the range 20--30\,\kms, with 10
pixels per resolution element. For further details about FUSE and
its on-orbit performance, see \citet{moos00} and \citet{sahnow00}.
The data were reduced using version 2.4.0 of the CALFUSE pipeline;
see \citet{fox05}\ and \citet{savage05a} for further details
regarding subsequent wavelength calibration steps and combination of
sub-integrations. (We note that over the 34 month span of the
observations, the quasar continuum level increased by about 25\%.)
In this study, we use primarily the LiF2A and SiC2A detector
segments that cover the wavelength ranges 1086.3--1182.0\,\AA\ and
916.5--1006\AA, respectively. In addition, the FUSE data are
rebinned to a sampling of about 3-4 bins per resolution element. The
CALFUSE pipeline and subsequent processing provides fully calibrated
(flux and wavelength) spectra that are accurate to 5\,\kms\ in
wavelength and about 10\% ($1-2 \times 10^{-15}$\,erg cm$^{-2}$\
s$^{-1}$\ \AA$^{-1}$) in flux. [Note that the FUSE spectra were
aligned with the STIS spectra using Galactic and intervening
absorption lines; see  \S\ref{sec:alignment}.] The long integration
time on this bright quasar provided signal-to-noise of 23 per
resolution element at 1150\,\AA\ (in the LiF2A detector segment) and
13 per resolution element at 950\,\AA\ (in the SiC2A detector
segment).

Selected portions of the FUSE LiF2A spectrum are show in
Figure~\ref{fig:fusedata}. The upper panel of the figure shows the
entire portion of the FUSE spectrum that is covered by the LiF2A
detector segment. For clarity, the spectra in this panel are shown
with a sampling of $\sim20$\,\kms, or 1 bin per FUSE resolution
element in this region of the spectrum. The underlying QSO continuum
from the power law fit to the STIS spectrum is also shown; the
contribution of the \ion{Ne}{8} and \ion{O}{4} broad emission lines
is readily apparent.

In the bottom panels, we show examples of the associated absorption
lines that are clearly detected in this spectrum. In these panels,
the spectra are shown with a sampling of $\sim$6\,\kms. The lower
left panel shows the portion of the spectrum covering the associated
absorption from the \ion{N}{4} $\lambda$765.148 and \ion{Ne}{8}
$\lambda$770.409 lines. In this panel, we also overlay our
Legendre-polynomial fit to the local continuum around the
\ion{Ne}{8} $\lambda$770.409 line to show that the detection of the
feature is robust. The line, which appears at an observed wavelength
of $\sim$1150\,\AA, is also readily apparent in the top panel. In
the bottom right panel, we show another region of the FUSE LiF2A
spectrum that covers the \ion{Ne}{8} $\lambda$780.324 and \ion{O}{4}
$\lambda$787.711 lines. The \ion{Ne}{8} $\lambda$780.324 line is not
detected in our spectrum down to a 3$\sigma$ equivalent width limit
of 103\,m\AA. The \ion{Ne}{8} $\lambda$770.409 line has an
equivalent width of 140 $\pm$ 22\,m\AA, so the non-detection of the
weaker line is consistent with a physically meaningful doublet ratio
(i.e, $1 \leq W_\lambda(\lambda770.409)/W_\lambda(\lambda780.324)
\leq 2$). We conclude that the detection of the absorption feature
at $\sim$1150\,\AA\ and its identification with the  \ion{Ne}{8}
$\lambda$770.409 line is reliable.

\subsection{A Remark on the Wavelength Cross-Calibration}
\label{sec:alignment}

\begin{figure*}
\epsscale{0.5}
\rotatebox{-90}{\plotone{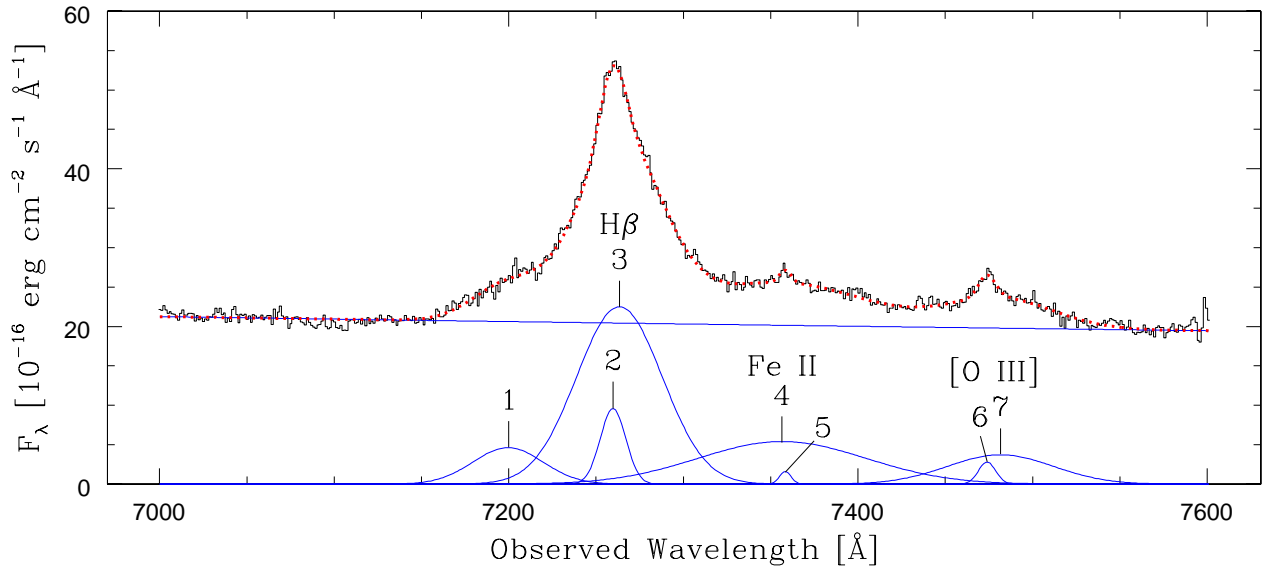}}
\protect\caption[Optical Emission Lines]{We show a portion of the
\citet{grupe99} optical spectrum of HE 0226-4110 (histogram) which
covers the H$\beta$\ and [\ion{O}{3}] $\lambda$5007 lines. The
smooth curves show the contributions to the best-fit from the
underlying power-law continuum and each emission line component. The
dotted curve shows the total best-fit spectrum. Each emission line
component is coded with a number that appears over the component
that is used to cross-reference with the fit parameters listed in
Table~\ref{tab:emission}. The locations of the H$\beta$,
\ion{Fe}{2}, and [\ion{O}{3}] emission lines are also labelled.}
\label{fig:emission}
\end{figure*}

Of particular importance to this study, the FUSE and STIS data were
aligned to a common (heliocentric) wavelength scale using the
Galactic \ion{C}{2} $\lambda\lambda$1334.532, 1036.337 and
\ion{Fe}{2} $\lambda\lambda$1144.938, 1608.451 lines, as well as the
Lyman $\gamma$\ and \ion{C}{3} $\lambda$977.020 IGM lines of the
$\zabs = 0.2070$\ absorber. The latter two lines appear in the
wavelength region 1149--1182\,\AA\ where the FUSE and STIS spectra
overlap. A portion of the overlapping region is shown in
Figure~\ref{fig:alignment}. The shift required to align the FUSE
spectra to the heliocentric velocity scale provided by the STIS
calibration was 3-5\,\kms, which is within the residual zero-point
uncertainty reported by \citet{fox05}. The alignment of these
spectra is important to this study, because there is an apparent
ionization-dependent velocity shift in the AAL at $\zabs=0.4925$\
between the {\ion{O}{3}} $\lambda832.927$\ line and the
\ion{O}{4}$\lambda787.711$\ line (see \S\ref{sec:aal}.1). The former
line appears in the STIS spectra, while the latter line appears in
the overlapping region and is flanked by two IGM lines from the same
system (at $\zabs = 0.2070$). The alignment of the IGM features
gives us confidence in the magnitude of the shift, and,
consequently, in the reality of the apparent velocity shift in the
associated absorber.

\subsection{Comments on \ion{Ne}{8}}

The detection of the high-ionization \ion{Ne}{8} $\lambda$770.409
line is fortuitous as it provides a further bridge between
lower-ionization associated \ion{C}{4} absorption and the
higher-ionization X-ray absorption by \ion{O}{7} and \ion{O}{8}
\citep[e.g.,][]{mew95,brandt}. To our knowledge, \ion{Ne}{8} has
only been detected in associated narrow absorption in three other
quasars, UM\,675 \citep{ham95}, 3C\,288.1 \citep{ham00},
HDFS-QSO\,J$2233-606$\ \citep{ps99}, and possibly HS\,1700+6416
\citep{pet96}. Previous detections of \ion{Ne}{8} absorption in
BALQSOs include PG 0946+301 \citep{arav99}, SBS\,1542+541
\citep{telfer98}, and Q\,0226-1024 \citep{korista92}. We note also
that the FUSE {\it composite} spectrum of \citet{scott04b} shows a
peculiar absorption feature just blueward of the
\ion{Ne}{8}+\ion{O}{4} emission at a rest wavelength of
$\sim$730\,\AA. [As noted by \citet{scott04b}, this feature was also
present, though at a much weaker level, in the rest-frame extreme
ultraviolet HST composite from \citet{telfer02}.] The HE\,0226-4110
FUSE spectrum presented here was used in that composite, but the
associated absorber does not contribute to that feature.

\section{Quasar Properties}
\label{sec:quasar}

Before proceeding with an analysis of the associated absorption at
$\zabs=0.4925$, we first discuss the general UV/optical, radio, and
X-ray properties of the quasar, as this is important to
understanding the environment of the absorbing gas.

\subsection{UV/Optical}
\label{sec:uvopt}

\input{tab1.tex}

The redshift of the quasar as measured by the peak of the
\ion{Mg}{2} emission line {\citep{rei96}} is $\zem = 0.495 \pm
0.001$. In general, it is more favorable in the study of associated
absorbers to use redshifts determined from narrow emission lines (as
these can be measured to higher precision) or from high-excitation
lines which are thought to arise from the quasar accretion disk (not
the outflow), and may more accurately reflect the systemic redshift.
To this end, we have analyzed the optical spectrum obtained by
\citet{grupe99} using the ESO Faint Object Spectrograph and Camera
($\approx$5\,\AA\ FWHM resolution with 1.2\,\AA/pixel sampling). The
spectrum covers the H$\beta$\ and [\ion{O}{3}] $\lambda$5007
emission lines and is shown in Figure~\ref{fig:emission}. (At the
redshifted position of the [\ion{O}{3}] $\lambda$5007 line, the
resolution of the spectrum is $\approx$200\,\kms\ with a sampling of
48\,\kms.) Since the H$\beta$\ and [\ion{O}{3}] $\lambda$5007
emission lines lie on top of an \ion{Fe}{2} multiplet, we model the
spectrum with an underlying power-law continuum with Gaussian
emission line components:
\begin{equation}
F_\lambda = F_{\lambda_{\mathrm{o}}} \left [ \left ( {{\lambda}
\over {\lambda_\mathrm{o}}} \right )^\beta + \sum_{i=1}^{m}
w_{\mathrm{i}} e^{- {1 \over 2} \left ( {{\lambda -
\lambda_{\mathrm{i}}} \over {\sigma_{\mathrm{i}}}} \right )^2}
\right ],
\end{equation}
where $\lambda_{\mathrm{o}}$\ is an arbitrary reference wavelength,
$F_{\lambda_{\mathrm{o}}}$\ is the normalizing continuum flux at
that wavelength, $\beta$\ is the continuum power-law index, $m$\ is
the number of emission line components, and $w_{\mathrm{i}}$,
$\lambda_{\mathrm{i}}$, and $\sigma_{\mathrm{i}}$\ are the relative
strength, central wavelength, and width of each emission line
component, respectively. We used a Marquardt-Levenburg least-squares
algorithm {\citep{nrpress}} to determine the best-fit for a given
number of emission components, and the F-test to choose the optimal
number of statistically significant components. In our fitting
algorithm, we use the entire spectrum and allow both power-law and
emission line parameters to vary simultaneously to minimize the
$\chi^2$\ (defined in the usual manner). The errors on all
parameters are given by the diagonal elements of the covariance
matrix (the inverse of the $\chi^2$\ curvature matrix). In
Figure~\ref{fig:emission} we also show the best-fit model, as well
as the separate contributions from the power-law continuum and each
emission-line component. The parameters of the fit are listed in
Table~\ref{tab:emission}. The fit appears reasonable, especially in
regions far from the emission lines, so we are confident in the
parameters of the underlying power-law.

Both the H$\beta$\ and [\ion{O}{3}] $\lambda$5007 emission lines are
well-described by the superposition of a broad component
(FWHM$=2861\pm191$\,\kms) at $\zem=0.4938\pm0.0001$\ and a narrow
component (FWHM$=966\pm86$\,\kms) at $\zem=0.4928\pm0.0001$. Three
additional components are required to yield a good fit to the data.
We identify one component with the \ion{Fe}{2} $\lambda$4924
multiplet. Since our goal is only to measure the redshift and
intensity of the [\ion{O}{3}] $\lambda$5007 line (to compare with
the associated absorption in the \ion{O}{3} $\lambda$832.927 line),
we do not attempt to further model the \ion{Fe}{2} emission.

In addition to the fit of the optical continuum and emission lines,
we have also fit the STIS spectrum which includes the Lyman limit of
the associated absorber. This is important for characterizing the
ionizing radiation field assumed later in our photoionization
models. Our method for fitting the UV spectrum was identical to that
of the optical spectrum (described above), and the underlying
power-law is shown in Figure~\ref{fig:stisdata}. The extrapolation
of this power-law to the FUSE band is also shown in
Figure~\ref{fig:fusedata}. In both figures, the fit appears quite
reasonable. The parameters of the power-law fit are:
$F_{\lambda_\mathrm{o}}=(1.74 \pm 0.01) \times 10^{-14}$\ erg
cm$^{-2}$\ s$^{-1}$\ \AA$^{-1}$, $\beta=-1.42 \pm 0.02$ at
$\lambda_\mathrm{o}=1395.55$\ \AA. We recast this fit into frequency
units ($F_\nu \propto \nu^{\alpha_{\mathrm{UV}}}$) and to the
rest-frame of the quasar: $F_{\nu_{\mathrm{o}}}=(5.00 \pm 0.03)
\times 10^{-27}$ erg cm$^{-2}$\ s$^{-1}$\ Hz$^{-1}$,
$\alpha_{\mathrm{UV}}= -0.58 \pm 0.02$\ at 1 Rydberg. As in
Table~\ref{tab:emission}, we adopt a luminosity distance of 2774.9
Mpc ($z=0.493$\ for a $\Omega_\Lambda=0.73$,
$\Omega_\mathrm{m}=0.27$ cosmology). Thus, the luminosity density of
the quasar at 1 Rydberg is $\log
L_\nu[\mathrm{erg~s^{-1}~Hz^{-1}}]=30.65$.

\subsection{Radio}

Another important quasar property to consider in understanding this
associated absorber is the radio-loudness (defined as the ratio of
flux densities at 5\,GHz and 2500\,\AA, $R^*$), since it appears
that absorption which is clearly related to quasar outflows (i.e.,
broad absorption lines) is connected to the presence and strength of
radio emission. The precise nature of this connection is not yet
clear. Strong, narrow, associated \ion{C}{4} absorption is
preferentially found in steep-spectrum, radio-loud ($R^*>10$)
quasars \citep[e.g.][]{foltz88,ves03}, while broad absorption lines
are predominantly found in radio-quiet quasars
\citep[e.g.][]{turn88b,weymann91,bro98,becker00}.

\begin{figure*}
\epsscale{1.1}
\plotone{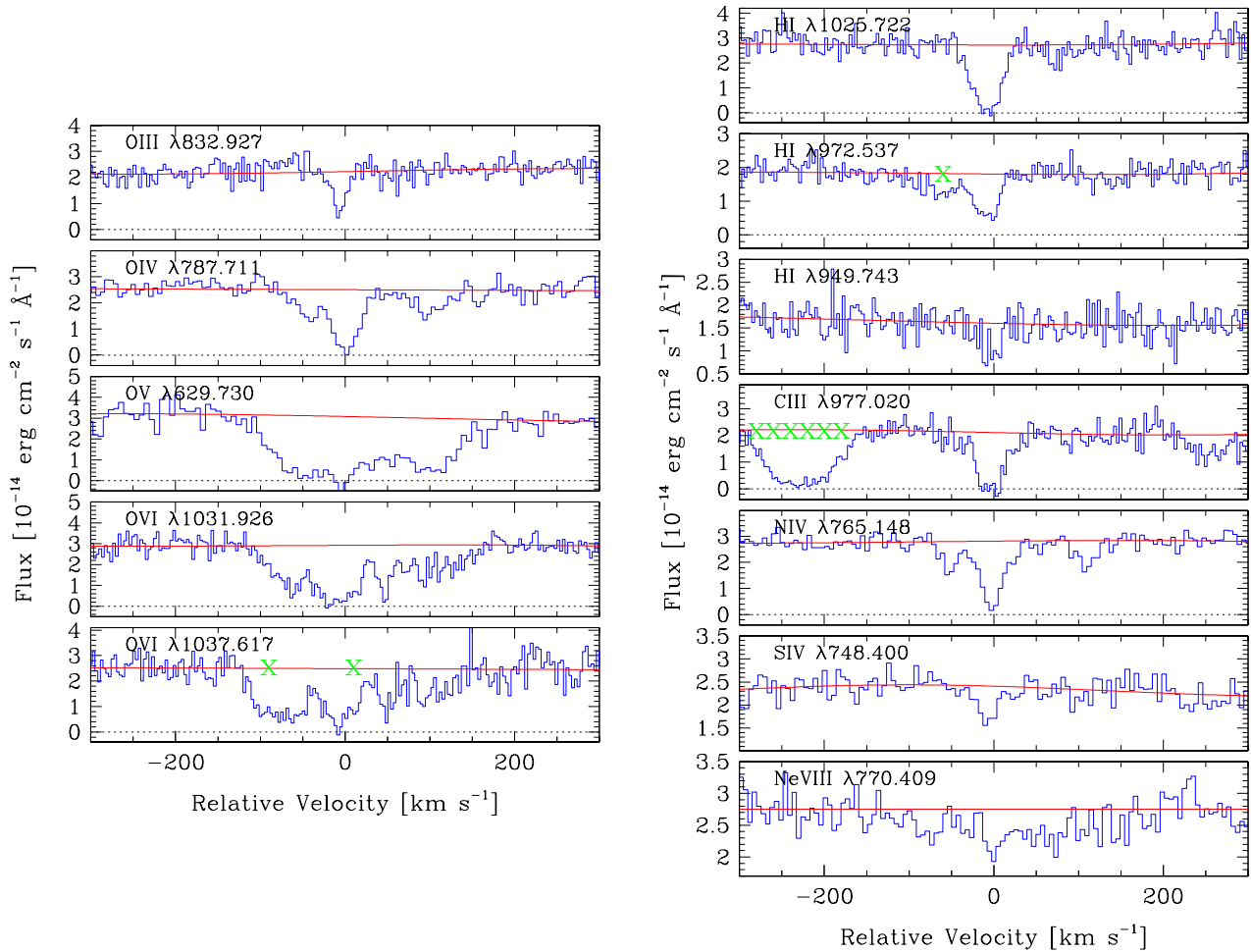}
\protect\caption[Velocity Stack]{Plots of flux versus velocity
(relative to $\zabs=0.4925$) of ions detected in the FUSE and STIS
spectra. Velocities where absorption is blended by a feature from an
unrelated system are marked with an `x'. The smooth curves are our
assessment of the local continuum using low-order Legendre
polynomials. On the left, we plot all detected transitions from
species of oxygen, which clearly show the ionization-dependent
differences in the kinematics. On the right, we show transitions
from all other detected species. Due to the shallowness of the
\ion{Ne}{8} $\lambda$770.409 absorption, the profile and continuum
fit are not well-represented in this expanded velocity range. We
refer the reader to the bottom left panel of
Figure~\ref{fig:fusedata} for a better view of this absorption
profile and its continuum fit.} \label{fig:velstack1}
\end{figure*}

This quasar is not detected by the Parkes-MIT-NRAO survey
{\citep{gw93}} down to a $4.4\sigma$\ flux density limit of 47.3 mJy
at 4.85 GHz. This limit corresponds to a radio-loudness of $\log R^*
< 1$, which is radio-quiet by the definition of {\citet{kel89}}. The
radio-loudness is an apparently important indicator of the soft
X-ray spectral slope \citep{laor97}, as well as being one of a
collection of properties (including the [\ion{O}{3}] $\lambda$5007
luminosity, and H$\beta$\ emission line full-width at half-maximum
intensity) that are apparently related among optically-selected
quasars \citep[that is, Eigenvector 1 from the principal components
analysis of][]{bg92}. The underlying connection among these
properties is generally interpreted as the accretion-rate of
material onto the black hole relative to Eddington,
$\dot{M}/\dot{M}_{\mathrm{edd}}$. The radio-loudness of this quasar
and its relatively weak [\ion{O}{3}] $\lambda$5007 emission indicate
that the accretion-rate relative to Eddington is low in comparison
with other quasars.

\subsection{Soft X-ray}

\begin{figure*}
\epsscale{1.1}
\plotone{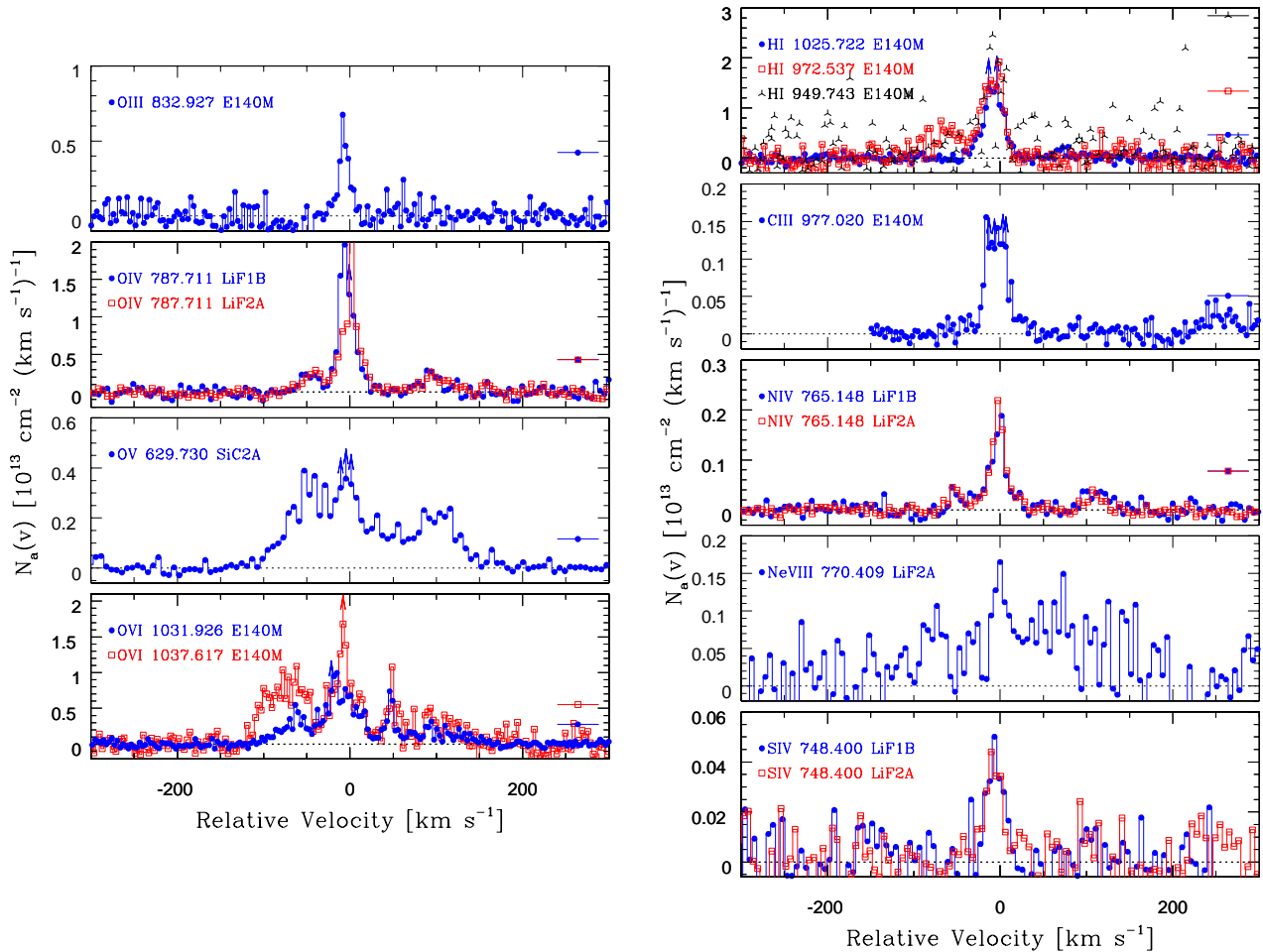}
\protect\caption[Apparent Column Density Profiles]{Apparent column
density profiles of the detected species as a function of velocity
relative to $\zabs=0.4925$. Transitions from a given species are
overplotted in the same panel. A horizontal bar on the right side of
each panel marks the apparent column density where each transition
reaches unit optical depth. FUSE data are shown with a sampling of 3
bins per resolution element, while STIS data are shown with a sampling
of 2 bins per resolution element. Species of oxygen are plotted on the
left, while other species are plotted on the right.}
\label{fig:oxygnaov}
\end{figure*}

There seems to be a connection between absorption in ultraviolet
transitions (in particular \ion{C}{4} $\lambda\lambda$
1548.204,1550.781) and ``warm absorbers'' in the soft X-rays (e.g.,
from \ion{O}{7}, and \ion{O}{8} edges) from both studies of
statistical coincidence {\citep[e.g.,][]{brandt,mon01}} and from
analyses of specific sight lines
{\citep[e.g.][]{scott04,cren03,ngc3783iii,ngc3783ii,mew95}}. Since
the absorption in these two bands results from species with widely
differing ionization potential, the precise causal connection is
unclear. HE\,0226-4110 was detected by the {\it ROSAT} All-Sky
Survey \citep[][RASS]{rassbsc} with a count rate of
$0.56\pm0.06$\,cps in the 0.2-2.0\,keV bandpass of the PSPC.
\citet{grupe98} fit the RASS PSPC X-ray spectrum of HE\,0226-4110
assuming the intrinsic spectrum is a single power-law and reported a
best-fit spectral index of $\alpha_{\mathrm{X}}=-2.1 \pm 0.0.2$\
with a total absorbing column density of $N$(H)$=(1.78 \pm 1.2)
\times 10^{20}$\, cm$^{-2}$. The origin of this total absorbing
column density can be attributed to Galactic absorption. [From a fit
to damped Lyman $\alpha$\ and $\beta$\ lines, \citet{fox05} find a
Galactic \ion{H}{1}\ column density of $1.2^{+0.4}_{-0.3} \times
10^{20}$\,cm$^{-2}$. Similarly, \citet{wakker03} report an
\ion{H}{1}\ column density of $1.55 \pm 0.01 \times
10^{20}$\,cm$^{-2}$\ from a fit to 21\,cm emission in this
direction.] \citet{grupe98} find no evidence for a significant
amount of the absorption in the soft- X-rays, and in particular from
\ion{O}{7} and \ion{O}{8}.

In further characterizing the ionizing radiation field, we compute
the two-point 2\,keV--to--2500\,\AA\ spectral index, $\aox$, which
provides a measure of the relative importance of the X-ray emission
to the UV emission. Using our power-law fit to the STIS spectrum,
and the \citet{grupe98} power-law fit to the ROSAT spectrum, we find
$\aox=-1.46 \pm 0.04$, which is close to the average value for
unabsorbed quasars {\citep[$\langle \aox \rangle \approx
-1.5$;][]{brandt}.

\section{Associated Absorber Properties at \\ $\mathbf{\zabs=0.4925}$}
\label{sec:aal}

\input{tab2.tex}

For associated absorption, our {\it HST} STIS and {\it FUSE} spectra
the spectra cover the rest-frame wavelength range 610--1140\,\AA. In
this range, we detect absorption at $\zabs = 0.4925$\ in \ion{H}{1}
Lyman $\beta$, $\gamma$, $\delta$, \ion{C}{3} $\lambda$977.020,
\ion{N}{4} $\lambda$765.148, \ion{O}{3} $\lambda$832.927, \ion{O}{4}
$\lambda$787.711, \ion{O}{5} $\lambda$629.730, \ion{O}{6}
$\lambda\lambda$1031.926, 1037.617, \ion{Ne}{8} $\lambda$770.409,
and \ion{S}{4} $\lambda$748.400. Other transitions, notably
\ion{C}{2} $\lambda$1036.337, \ion{O}{1} $\lambda$877.879, and
\ion{O}{2} $\lambda\lambda$834.466, 833.329 are also covered in our
spectrum, but are not detected. \ion{O}{3} $\lambda$702.332 is also
covered by our spectrum and has a line strength which is larger than
the 832.927\,\AA\ transition, but it is blended with Galactic
\ion{Ar}{1} $\lambda$1048.220. [For this study, we adopt transition
wavelengths and oscillator strengths from \citet{morton03} for
transitions above 912\,\AA, and from \cite{vbt94} for transitions
below 912\,\AA.] In Figure~\ref{fig:velstack1}, we plot the detected
transitions (histogram) in velocity with $\zabs=0.4925$\ defining
the velocity zero-point. (Integration of the \ion{O}{3}
$\lambda$832.927 line yields a centroid redshift of $\zabs=0.49246
\pm 0.00006$. Henceforth, all quoted velocities are referenced to
$\zabs=0.4925$.) The smooth curves overplotted on the data indicate
our estimated continua obtained by fitting a low-order ($\leq$5)
Legendre polynomial to the adjoining continuum regions. These
continuum fits follow the formalism described by \citet{ss92}.

We make two remarks regarding the continuum fits to the \ion{O}{3}
$\lambda$832.927 and \ion{Ne}{8} $\lambda$770.409 lines. Inspection
of the continuum fit to the \ion{O}{3} $\lambda$832.927 lines shows
several bins blueward of the absorption profile (corresponding to
the velocity range $-100 \leq v$[\kms]$ \leq -50$) that appear above
the continuum. In order to fit these bins and lower them to the
continuum level, we would require an polynomial order of at least 7,
larger than what is prescribed by \citet{ss92}. Using such a high
order, we would be in danger of fitting out some of the absorption.
Outside of the velocity range $-100 \leq v$[\kms]$ \leq +5$\ the fit
appears quite good. In addition, the fit to the broad and shallow
trough of the \ion{Ne}{8} $\lambda$770.409 is not well-represented
in the expand view of Figure~\ref{fig:velstack1}. We refer the
reader to the lower left panel of Figure~\ref{fig:fusedata} where we
show a larger wavelength range with the Legendre-polynomial fit to
the local continuum superimposed. The continuum placement is
reasonable.

In order to remove biases due to transition strengths ($\log
f\lambda$) and compare profiles of species on a common (and
physically meaningful) scale, we transform the flux profiles of each
detected transition into apparent column density profiles using the
method described by {\citet{ss91}}.
\begin{equation}
\acd(v) = {{m_{\mathrm{e}} c} \over {\pi e^2}} {1 \over
{f\lambda_{\circ}}} \ln \left [ {{I_{\mathrm{c}}(v)} \over {I(v)}}
\right ], \label{eq:naov}
\end{equation}
where $f$\ and $\lambda$\ are the oscillator strength and rest
wavelength of the transition, and {$I_{\mathrm{c}}(v)$} and {$I(v)$}
are the continuum and attenuated fluxes, respectively. These are
shown in Figure~\ref{fig:oxygnaov}. Apparent column density profiles
from multiple transitions from a given species are useful in testing
for unresolved saturated structure, blends with transitions of other
species from other absorption-line systems, and for assessing the
levels of unocculted flux (e.g., scattered light within the
instrument, or elevated background levels). For systems that are
potentially intrinsic to the quasar, apparent column density
profiles provide a means of assessing the fraction(s) of the quasar
continuum/emission line region that is occulted by the absorbing gas
\citep[e.g.,][]{scott04,gan99,ham97a,bs97}.

For the associated absorption along this sight line, our spectra
cover multiple transitions of \ion{H}{1} (Lyman $\beta$\ and higher
order Lyman series), and \ion{O}{6} (the resonant doublet
$\lambda\lambda$ 1031.927,1037.617). In Figure~\ref{fig:oxygnaov},
we overlay the apparent column density profiles of the \ion{O}{6}
doublet (left side, bottom panel). The apparent excess of column
density in the \ion{O}{6} $\lambda$1037.617 transition over the
relative velocity range $-120 \leq v$[\kms]$\leq 0$\ is due to
Galactic and high velocity cloud absorption by the \ion{C}{4}
$\lambda$1548.204 line \citep{fox05}. At all other velocities, the
\ion{O}{6} $\lambda$1031.927 and $\lambda$1037.617 lines are in
excellent agreement (within the errors) and there is no evidence for
unresolved saturated structure, or for partial coverage of the
quasar continuum/broad emission line region. Also, strongly
saturated lines such as \ion{C}{3} $\lambda$977.020 are black in the
line core (i.e., no flux is significantly detected), which further
indicates that the absorber fully covers the continuum flux source.

\begin{figure*}
\epsscale{0.8}
\rotatebox{-90}{\plotone{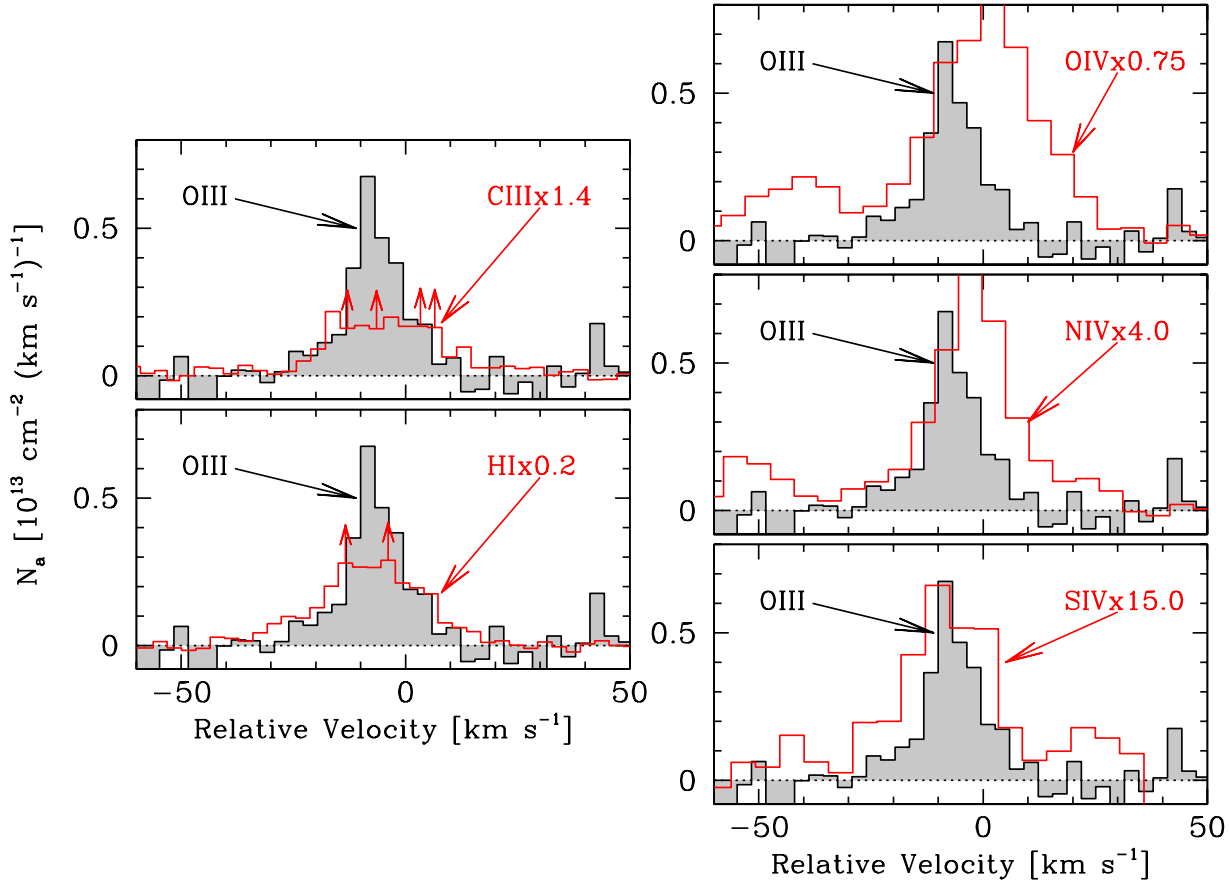}}
\protect\caption[Comparison of O III to other species]{In the above
panels, we overlay the \ion{O}{3} apparent column density profiles
(shaded histogram) on top of the \ion{H}{1}, \ion{C}{3}, \ion{N}{4},
\ion{O}{4}, and \ion{S}{4} apparent column density profile. Vertical
arrows are shown for pixels where the flux is negative due to Poisson
fluctuations at low flux levels. The \ion{H}{1} and \ion{C}{3}
profiles are scaled to match the
\ion{O}{3} profile in unsaturated regions. The \ion{N}{4} and
\ion{O}{4} profiles are scaled to match the blue wing of the
\ion{O}{3} profile. The \ion{S}{4} profile is scaled to match the
peak of the \ion{O}{3} profile.} \label{fig:narrow1a}
\end{figure*}

There is a weak intervening IGM Ly $\alpha$\ line at $\zabs=0.19374$
\citep{lehner05} that is blended with the Ly $\gamma$\ line from the
associated absorber at $v\sim -65$\,\kms. The intervening Ly
$\alpha$ line appears somewhat broad and shallow, and the Ly
$\gamma$\ line of the associated absorber lies on its wing. Due the
noisiness of the data on the blue wing of the z=0.19374 Ly $\alpha$
line, it is difficult to assess its contribution to the z=0.4925 Ly
$\gamma$ profile, and the level of saturation in the core of the
Lyman $\beta$. (The Ly $\delta$ line is detected, but also too noisy
for this purpose.) We return to this issue in \S\ref{sec:narrow} by
using the \ion{O}{3} $\lambda$832.926 profile. We note here that
\citet{lehner05} report $\log
N$(\ion{H}{1})[cm$^{-2}$]$=13.20\pm0.06$\ and
$b$(\ion{H}{1})$=28.7\pm6.0$\,\kms\ for the intervening
$\zabs=0.19374$\ Ly $\alpha$\ line.

We make one final note regarding the apparent column density
profiles as a whole before continuing with more detailed
investigations of the various components. The rest-frame transitions
in the 732--792\,\AA\ wavelength range lie in both the LiF1B and
LiF2A detector segments of the {\it FUSE} spectra. We overlay the
apparent column density profiles from both channels for the
\ion{S}{4} $\lambda$748.400 (Figure~\ref{fig:oxygnaov}, right side,
bottom panel), \ion{N}{4} $\lambda$765.148
(Figure~\ref{fig:oxygnaov}, right side, middle panel), and
\ion{O}{4} $\lambda$787.711 (Figure~\ref{fig:oxygnaov}, left side,
second panel from the top) lines. The agreement between the LiF1B
and LiF2A profiles for \ion{N}{4} and \ion{S}{4} is excellent. For
the \ion{O}{4} profiles, there is a discrepancy in the bins near
0\,\kms\ where the line profile reaches zero flux. The profiles are
in excellent agreement otherwise, and we have no reason to suspect
either one. We attribute the discrepancy near 0\,\kms\ to Poisson
errors, and use the profiles from the LiF2A detector segment for our
investigation below (due the higher signal-to-noise). The excellent
agreement evident in Figure~\ref{fig:oxygnaov} indicates that the
LiF1 and LiF2 channels have very similar spectral resolutions.

\subsection{Narrow Components}
\label{sec:narrow}

\subsubsection{Low and moderate ionization gas at $v=-8$\,\kms\ (in the z = 0.4925 rest-frame)}

In Figures~\ref{fig:velstack1}--\ref{fig:oxygnaov}, absorption is
clearly detected at -8\,\kms\ in a wide variety of ionization
species, from neutral \ion{H}{1}, to low-ionization \ion{C}{3}\ and
\ion{O}{3}, to moderate-ionization \ion{N}{4}, \ion{O}{4},
\ion{S}{4}, to high-ionization \ion{O}{5}, \ion{O}{6}\ and
\ion{Ne}{8}. The kinematics of the absorption at this velocity is
complex (and possibly multi-phase). Neutral (\ion{H}{1}) and
low-ionization species (\ion{C}{3}, \ion{O}{3}) appear as a discrete
feature over the velocity range $-30 \leq v$[\kms]$ \leq +5$, while
the higher-ionization species (\ion{N}{4}, \ion{O}{4}) appear as a
broader feature over the velocity range $-30 \leq v$[\kms]$ \leq
+30$. The more highly-ionized species (\ion{O}{5-VI}, \ion{Ne}{8})
appear absorbed with little substructure over the entire velocity
range $-30 \leq v$[\kms]$ \leq +30$, not as a single discrete
feature. We focus here on the discrete features appearing in the
\ion{H}{1}, \ion{C}{3}, \ion{N}{4}, \ion{O}{3-IV}, and \ion{S}{4}
ions.

To investigate the number of phases that are required to explain the
absorption at -8\,\kms\ and the ionization conditions of each phase,
we first overlay the apparent column density profiles of the
neutral, low- and moderate-ionization species in
Figure~\ref{fig:narrow1a}. In the left panels of
Figure~\ref{fig:narrow1a}, we overplot the \ion{H}{1}\ and
\ion{C}{3}\ apparent column density profiles on top of the
\ion{O}{3}\ profile. For \ion{H}{1}, we use the Ly $\beta$ profile
because it provides the best and cleanest information in the shapes
of the wings without suffering from blends (as in the case of Ly
$\gamma$) or noise (as in the case of Ly $\delta$). We comment on
this further below. The kinematics of these species is remarkably
similar and suggest an origin in the same phase. (Note the gradual
trend toward increasing column density in the velocity range $-30
\leq v \leq -15$\,\kms, and the sharp drop-off at 5\,\kms.) The
reason for this asymmetry is unclear. If the absorption results from
the superposition of symmetric (e.g., Gaussian) profiles, then at
least two such components are needed to reproduce the observed
profile. However, it is not clear that such an assumption is
required \citep*[e.g.,][]{gan05a, akd02}. Regardless, if the
kinematics of the \ion{H}{1}\ and \ion{C}{3} in the velocity range
$-15 \leq v \leq +5$\,\kms\ are also well-traced by the \ion{O}{3}\
profile, we can use the \ion{O}{3}\ profile as a template to
estimate the level of saturation, and, more importantly, to obtain
column density ratios to constrain the ionization conditions of the
low-ionization phase. We summarize these ratios below.

In the right panels of Figure~\ref{fig:narrow1a}, we overplot the
\ion{O}{4}, \ion{N}{4}, and \ion{S}{4}\ profiles on top the
\ion{O}{3}\ profile. These overlays also show that the kinematics of
these profiles do not match that of the \ion{O}{3} profile. They are
wider, possibly resulting from a blend of different kinematic
components (see \S\ref{sec:narrow2}. With the simple assumption that
any \ion{O}{4}, \ion{N}{4} and \ion{S}{4} that co-exists in the
\ion{O}{3}-bearing gas have the same kinematic profile, we can place
limits on how much of these higher-ionization species relative to
\ion{O}{3} could arise in that phase (in the interest of providing
the most information to constrain ionization models). (Although
there is no apparent shift in the centroid of \ion{S}{4}
$\lambda$784.400 line relative to the \ion{O}{3} $\lambda$832.927
line, the kinematics of the two profiles do not match. While it is
possible that this could be an effect of the resolution, since the
\ion{S}{4} $\lambda$784.400 line appears in the FUSE LiF2 channel,
the profile spans over three resolution elements. Thus, we report
only a limit on the $N$(\ion{S}{4})$/N$(\ion{O}{3}) ratio.) In
Figure~\ref{fig:narrow1a}, we use the \ion{O}{3} profile, which
appears over the velocity range $-30 \leq v$[\kms]$ \leq +5$, to
estimate the following constraints on the column density ratios for
the \ion{O}{3}-bearing gas:
\begin{eqnarray}
\label{eq:narrow1}
N(\mathrm{H\,I})/N(\mathrm{O\,III}) \approx 5 \\
N(\mathrm{C\,III})/N(\mathrm{O\,III}) \approx 0.74 \nonumber \\
N(\mathrm{N\,IV})/N(\mathrm{O\,III}) \lesssim 0.25 \nonumber \\
N(\mathrm{O\,IV})/N(\mathrm{O\,III}) \lesssim 1.3 \nonumber \\
N(\mathrm{S\,IV})/N(\mathrm{O\,III}) \lesssim 0.07 \nonumber
\end{eqnarray}
In columns 4 and 5 of Table~\ref{tab:narrow}, we report our
equivalent width measurements and further summarize our column
density assessments of the \ion{O}{3}-bearing gas using the above
ratios. Direct integration of the \ion{O}{3}\ profile over the above
velocity range yields a column density of
$\acd$(\ion{O}{3})$=(8.4\pm1.1_{-1.7}^{+0.5}) \times 10^{13}$
cm$^{-2}$, where the first error is the {$1\sigma$} confidence error
resulting from statistical and continuum placement uncertainties and
the latter error is the systematic uncertainty resulting from the
choice of integration range. (To assess this latter error, we added
and subtracted 10\,\kms\ to the integration range.) We note that the
resulting \ion{H}{1}\ column density implied by eq.~\ref{eq:narrow1}
[$N$(\ion{H}{1})$ \approx (4.2\pm0.6_{-0.9}^{+0.3}) \times 10^{14}
$\,cm$^{-2}$] is consistent with the curve of growth of the Lyman
series equivalent widths. In addition, we integrated the region of
the spectrum where the \ion{O}{2} $\lambda$833.329 line ($\log
f\lambda=2.097$) is expected and find a limiting column density of
$N$(\ion{O}{2})$<1.3 \times 10^{13} \mathrm{cm}^{-2} (3\sigma)$\ for
this low-ionization phase.

\begin{figure}
\epsscale{1.1}
\rotatebox{0}{\plotone{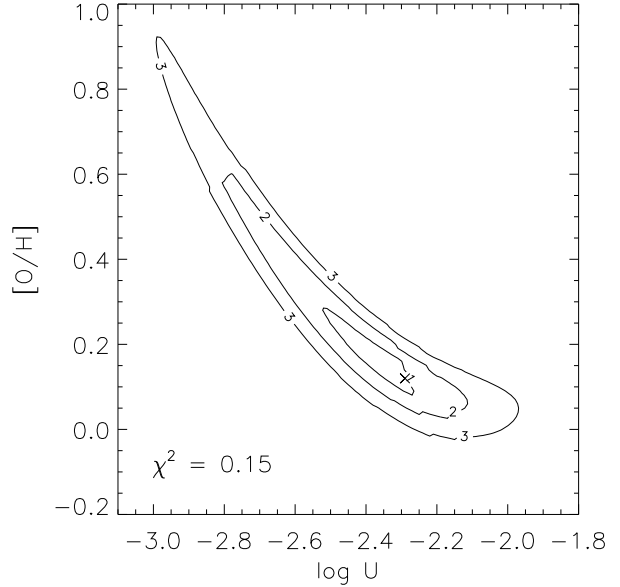}}
\protect\caption[Photoionization of $v\sim0$\,\kms\ component]{In
the plot we show $\Delta\chi^2$\ contours on the metallicity ([O/H])
- ionization parameter ($U$) plane for the low-ionization phase of
the $v=-8$\,\kms\ component probing the narrow emission line region.
The optimal parameters are marked with a cross while contours are
drawn for 1, 2, and $3\sigma$\ confidence regions.}
\label{fig:narrow1phot1}
\end{figure}

To understand the ionization structure of the \ion{O}{3}-bearing
phase, we assume that the absorption arises from a photoionized
plane-parallel slab with a solar relative abundance pattern
[$n$(C)$/n$(O) $=0.537$; \citet*{ags05}, and references therein] and
use the above column density ratios to infer the ionization
parameter, $U$, defined as the number density of ionizing photons
per baryon, and metallicity, [O/H]. We use the \cloudy\
photoionization code \citep{hazy} to compute the best-fit model,
specifying the radiation field as an AGN spectrum with the following
parameters (as described in \S\ref{sec:quasar}): $T=15,000$\,K,
$\alpha_{\mathrm{ox}}=-1.46$, $\alpha_{\mathrm{UV}}=-0.58$,
$\alpha_{\mathrm{x}}=-2.1$, normalized to $\log
L_\nu(1~\mathrm{Ryd})=30.65$. (See \S\ref{sec:quasar} for a
justification of these parameters.) In
Figure~\ref{fig:narrow1phot1}, we show $\chi^2$\ contours on the
[O/H] - $U$\ plane. The best fitting parameters are:
\begin{eqnarray}
\mathrm{[O/H]}=+0.12_{-0.03}^{+0.16}, \label{eq:narrow1phot}
\\
\log U=-2.29_{-0.23}^{+0.02}, \nonumber \\
\log N(\mathrm{H})=17.54_{-0.25}^{+0.04}. \nonumber
\end{eqnarray}
The total hydrogen column density of the \ion{O}{3}-bearing phase
results from a combination of the integrated \ion{O}{3} column
density (Table~\ref{tab:narrow}), the \ion{H}{1}/\ion{O}{3}\ column
density ratio from equation~\ref{eq:narrow1}, and the \ion{H}{1} ion
fraction implied by the ionization parameter:
$N$(H)$=N$(\ion{O}{3})$\times [N$(\ion{H}{1})$/N$(\ion{O}{3})$]
\times f_{\mathrm{H}{1}}^{-1}(U)$. The ionization parameter is most
tightly constrained by the \ion{O}{4}/\ion{O}{3} limit and
\ion{C}{3}/\ion{O}{3} (with the assumed solar relative C/O
abundance) ratio.

Increasing the carbon to oxygen relative abundance to [C/O]$=+0.1$\
yields a slightly better fit to the observations (best parameters:
[O/H]$\sim +0.45$, $\log U \sim -2.73$), since the lower constraint
on the ionization parameter provided by \ion{C}{3}/\ion{O}{3} is
relaxed. The best fit model presented above yields
\ion{O}{4}/\ion{O}{3}\ at the observational limit, while the
[C/O]$=+0.1$\ best-fit model predicts \ion{O}{4}/\ion{O}{3}\ well
below the limit. (The \cloudy\ model produces a factor of 1.3 less
\ion{N}{4} column density than the observational limit.) All other
best-fit models with non-solar carbon to oxygen abundances
([C/O]$>0.1$, [C/O]$<0$) fail to reconcile the \ion{O}{3}\ column
density and \ion{C}{3}/\ion{O}{3}\ column density ratio.
(Photoionization models with [C/O]$>0.1$\ predict ratios that are
too large; those with [C/O]$<0$ predict ratios that are too small.)
Such an enhancement may be reasonable given that the Fe/$\alpha$\
relative abundance appears to be enhanced in the emission-line
regions of AGN \citep[e.g.,][and references therein]{hf99}. There is
evidence from the atmospheres of Galactic halo stars
\citep[e.g.,][]{mcwilliam97} that the C/Fe is roughly constant (at
the solar value) over a large range in metallicity. If this
insensitivity can be extrapolated to the super-solar metallicities
of AGN line-emitting regions, then a slightly super-solar C/O
relative abundance in this $\zabs\sim0.5$\ absorber is reasonable.


We note here that the ionization fractions of \ion{O}{5} and
\ion{O}{6} for the \ion{O}{3}-bearing gas are $6.75 \times 10^{-2}$\
and $2.52 \times 10^{-3}$, respectively, compared to 0.4 for
\ion{O}{3}. This implies column densities of $N$(\ion{O}{5})$\sim
10^{13}$\,cm$^{-2}$, and $N$(\ion{O}{6})$\sim 5 \times
10^{11}$\,cm$^{-2}$. Integration of the \ion{O}{5} and \ion{O}{6}
over the velocity range $-30 \leq v$[\kms]$ \leq +5$\ yield apparent
column densities of $\acd$(\ion{O}{5})$=(1.1 \pm 0.1) \times
10^{14}$\,cm$^{-2}$, and $\acd$(\ion{O}{6})$=(2.4 \pm 0.4) \times
10^{14}$\,cm$^{-2}$. These high-ionization species must arise in a
phase/component of gas that is separate from the \ion{O}{3}-bearing
gas.

\subsubsection{Moderate and High ionization gas at $v=0$\,\kms\ (in the z = 0.4925 rest-frame)}
\label{sec:narrow2}

\begin{figure}
\epsscale{1.0}
\plotone{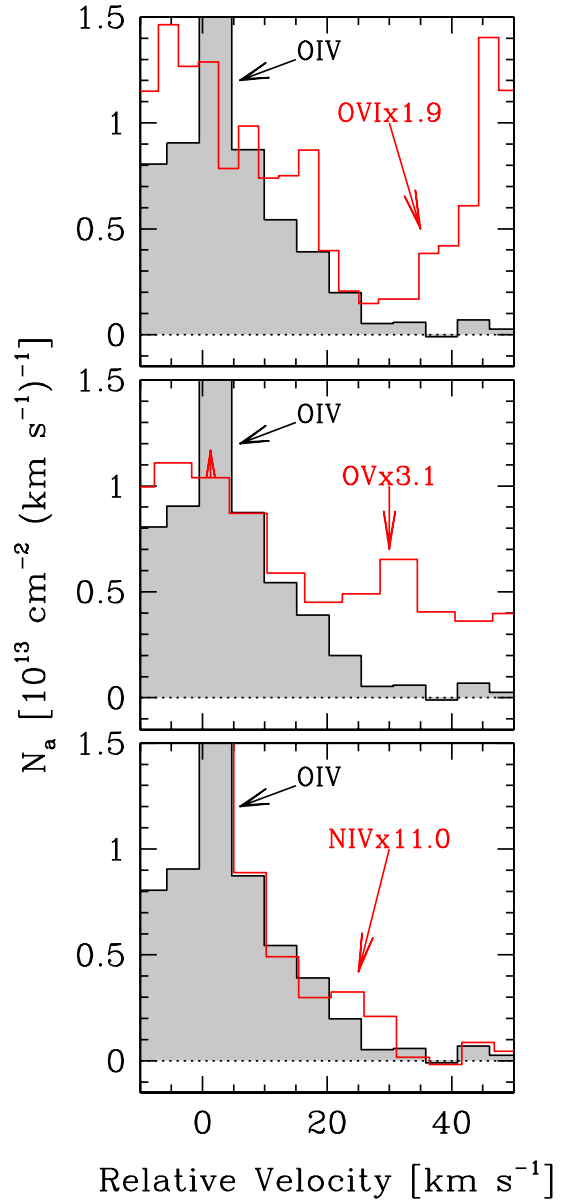}
\protect\caption[Comparison of O IV to other species]{In the figure,
we overlay the apparent column density profile of \ion{O}{4}\
(shaded histogram) on top of the \ion{O}{6}\ (top panel),
\ion{O}{5}\ (middle panel), and \ion{N}{4}\ (bottom panel) profiles
(unshaded histograms) in the region of the $v=0$\,\kms\ component.
The \ion{N}{4} profile has been scaled to match the \ion{O}{4}
profile in the velocity range $+5 \leq v$[\kms]$ \leq +30$. The
\ion{O}{5-VI}\ profiles have been scaled to the minimum allowable
value to hide the \ion{O}{4}\ profile in that same velocity range.}
\label{fig:narrow1b}
\end{figure}

As we pointed out in the previous section from the overlays of the
\ion{O}{4}, \ion{N}{4}, and \ion{S}{4} apparent column density
profiles on top of the \ion{O}{3} profile, we can only place limits
on the amount of higher-ionization species that co-exists in the
\ion{O}{3}-bearing gas. We return to the issue of the
higher-ionization species here. These higher-ionization species
appear over the velocity range $-30 \leq v$[\kms]$ \leq +30$. In
columns 6 and 7 of Table~\ref{tab:narrow} we report equivalent width
and apparent column density integrations of these and
high-ionization species (\ion{O}{5-VI}, \ion{Ne}{8}) over this
velocity range. (We do not report additional measurements of
\ion{H}{1}, \ion{C}{3}, or \ion{O}{3} as there is no additional
information over that reported in columns 4 and 5.)

As we pointed out earlier, there are differences in the apparent
velocity widths of the \ion{O}{4},\ion{N}{4}, and \ion{S}{4}
compared to \ion{O}{3}, \ion{H}{1}, and \ion{C}{3}. There are two
additional aspects of Figure~\ref{fig:narrow1a} that merit
consideration. First, there is a $\sim +8$\,\kms\ velocity shift in
the peak of the \ion{O}{4}\ and \ion{N}{4}\ profiles relative to the
\ion{O}{3}\ profile. Second, in the velocity range $+5 \leq v \leq
+30$\,\kms, there is significant column density apparent in
\ion{O}{4}, whereas none exists in the \ion{O}{3}, \ion{H}{1}, and
\ion{C}{3} profiles.

From this, we conclude that there is at least one additional
higher-ionization phase of gas at these velocities which produces
\ion{O}{4}, \ion{N}{4}, and possibly some \ion{O}{5}, and
\ion{O}{6}. This ``component'' is not necessarily restricted to the
velocity range $+5 \leq v$[\kms]$ \leq +30$, however. It is
certainly plausible that the absorption by this gas extends down to
velocities where the \ion{O}{3}-bearing gas appears. (For this
reason, we can only quote limits on the column density ratios in the
previous section.) In fact, as mentioned in the previous section,
additional gas in the velocity range common with that of the
\ion{O}{3}-bearing gas (i.e., $-30 \leq v$[\kms]$ \leq +5$) is
required to explain the column densities of the high-ionization
species \ion{O}{5-VI}.

Since much of the \ion{O}{4} and \ion{N}{4} column density in the
velocity range $-30 \leq v$[\kms]$ \leq +5$\ can be explained by the
\ion{O}{3}-bearing gas, we focus here on the remaining velocity
range $+5 \leq v$[\kms]$ \leq +30$\ (with the aforementioned
caveats). In Figure~\ref{fig:narrow1b}, we compare the apparent
column density profiles of \ion{O}{5-VI} and \ion{N}{4} to that of
\ion{O}{4}. In the velocity range $+5 \leq v$[\kms]$ \leq +30$,
there is good agreement between the kinematics of \ion{O}{4} and
\ion{N}{4}, so the two likely co-exist in the same phase of gas.
[Note that the \ion{N}{4} profile has been scaled vertically.]
Figure~\ref{fig:narrow1b} also shows that the kinematics of
\ion{O}{5-VI} do not match \ion{O}{4} in the velocity range $+5 \leq
v$[\kms]$ \leq +30$. This implies that the absorbing gas of those
two species at these velocities arise in a separate phase/component
(probably one of the broad components described in
\S\ref{sec:broad}). Using the comparison of the apparent column
density profiles we can still place limits on how much \ion{O}{5-VI}
could arise in the \ion{O}{4}-bearing gas at these velocities.
[Here, we again make the simple assumption that any \ion{O}{5-VI}
that co-exists in the \ion{O}{4}-bearing gas must have the same
kinematics as the \ion{O}{4} over those velocities.] We find the
following constraints:
\begin{eqnarray}
N(\mathrm{N\,IV})/N(\mathrm{O\,IV}) \approx 0.1 \\
N(\mathrm{O\,V})/N(\mathrm{O\,IV}) \lesssim 0.32 \nonumber \\
N(\mathrm{O\,VI})/N(\mathrm{O\,IV}) \lesssim 0.53 \nonumber
\end{eqnarray}

For the \ion{O}{4}-bearing gas in the velocity range $+5 \leq
v$[\kms]$ \leq +30$, we find
$\acd$(\ion{N}{4})/$\acd$(\ion{O}{4})$\approx 0.1$\ (compared to
$\approx 0.2$\ for the \cloudy\ model of the \ion{O}{3}-bearing
gas). In the photoionization models described above, the smallest
achievable $N$(\ion{N}{4})/$N$(\ion{O}{4}) ratio is $\approx 0.14$,
which occurs at an ionization parameter of $\log U\approx-2.6$. This
is problematic for two reasons. First, at this ionization level, we
would expect to see \ion{O}{3} at these velocities which we do not
(see Figures~\ref{fig:oxygnaov} and \ref{fig:narrow1a}) down to a
$3\sigma$ confidence limit of $1.6 \times 10^{13}$\,cm$^{-2}$.
Second, while the ratio is close, this ionization level does not
really explain the observed ratio. There are two possible
conclusions from this simple analysis. Either pure photoionization
by the quasar may be the only ionization mechanism for the
\ion{O}{4}/\ion{N}{4}-bearing gas at these velocities, or the N/O
 relative abundance deviates from our assumed solar ratio
\citep[N/O=0.132,][]{ags05}.

If the ionization mechanism of this gas is predominantly
photoionization by the quasar, then the lack of \ion{O}{3} implies
that the ionization parameter must be larger than $\log U \approx
-1.7$. At this ionization level, and with solar relative abundances,
the \ion{N}{4}/\ion{O}{4} ratio is $\approx 0.17$. Thus, the N/O
abundance must be lowered by at least $\approx 0.07$\ dex relative
to the solar value (i.e, [N/O]$\lesssim -0.07$) to match the
observed \ion{N}{4}/\ion{O}{4} ratio.


\subsubsection{High-ionization gas at $v=+50$\,\kms\ (in the z = 0.4925 rest-frame)}
\label{sec:narrow3}

\begin{figure}
\epsscale{1.0}
\plotone{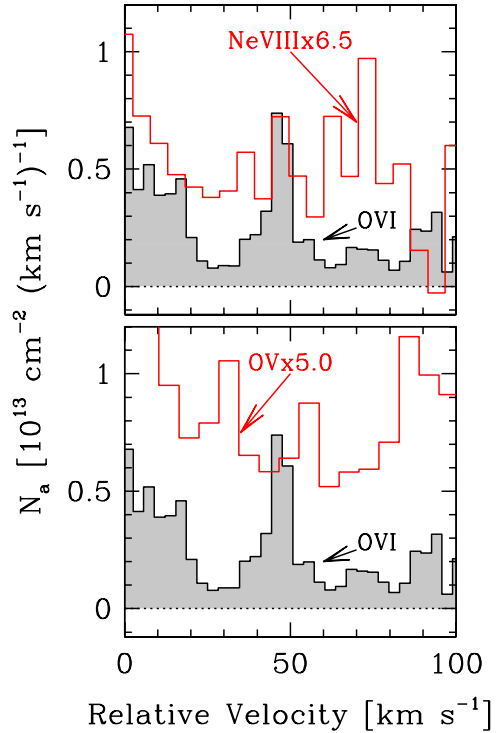}
\protect\caption[Comparison of O VI to other species]{In the figure,
we overlay the apparent column density profile of \ion{O}{6}\
(shaded histogram) on top of the \ion{O}{5}\ (bottom panel) and
\ion{Ne}{8}\ (top panel) profiles (unshaded histograms) in the
region of the $v=+50$\,\kms\ component. The \ion{O}{5}\ and
\ion{Ne}{8}\ profiles have been scaled to the minimum allowable
value to hide the \ion{O}{6}\ profile.} \label{fig:narrow2}
\end{figure}

In Figures~\ref{fig:velstack1}-\ref{fig:oxygnaov}, there is clearly
a narrow component detected in \ion{O}{6}\ at +50\,\kms\ in the
$z=0.4925$\ rest-frame. This component is not detected in any low-
to moderate-ionization species (i.e., \ion{H}{1}, \ion{C}{3},
\ion{N}{4}, \ion{O}{3}, \ion{O}{4}, \ion{S}{4}). Absorption is
detected at this velocity in \ion{O}{5}\ and \ion{Ne}{8}. However,
it appears that those species arise from the broader component at
$v\sim100$\,\kms. If so, the only ion detected in this narrow
component is {\ion{O}{6}}. To place constraints on the column
densities of species (for the consideration of ionization models
below), we integrate the apparent column density profiles of
non-detected species and \ion{O}{6} over the velocity range $+30
\leq v \leq +60$\,\kms. For \ion{O}{5}\ and \ion{Ne}{8}, we take the
same approach as in the previous section and consider how much of
the absorption observed in this velocity range could be attributed
to the \ion{O}{6}\ component. In Figure~\ref{fig:narrow2}, we plot
the \ion{O}{6} apparent column density profile over those of
\ion{O}{5}\ (bottom panel) and \ion{Ne}{8}\ (top panel). In each
case, we have scaled the apparent column density profile to match
the level of the \ion{O}{6}. This yields the following constraints:
\begin{eqnarray}
N(\mathrm{O\,V})/N(\mathrm{O\,VI}) \lesssim 0.2 \\
N(\mathrm{Ne\,VIII})/N(\mathrm{O\,VI}) \lesssim 0.15 \nonumber
\end{eqnarray}
In columns 8 and 9 of Table~\ref{tab:narrow}, we summarize our
equivalent width measurements (over the velocity range $+30 \leq
v$[\kms]$ \leq +60$) and column density constraints.

\begin{figure}
\epsscale{1.0}
\rotatebox{-90}{\plotone{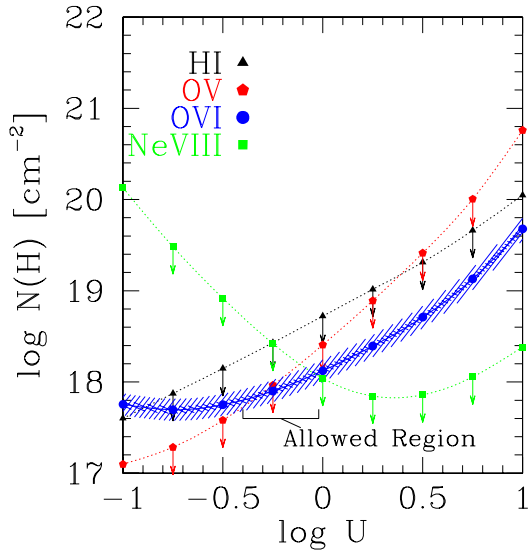}}
\protect\caption[Phtoionization of $v\sim50$\,\kms\ component]{The
figure shows isopleths of species column density in total column
density [$N$(H)]- ionization parameter ($U$) space for the $v =
+50$\,\kms\ component. The relevant species for each curve is
distinguished by the symbols shown in the corner. For \ion{O}{6},
the hatched regions around the curve indicate the $1\sigma$\ and
$3\sigma$\ confidence uncertainties. For all other species, the
curves are shown as upper limits. The region of parameter space that
satisfies all column density information is marked ``Allowed
region.'' The models assumed a metallicity [O/H]$=+0.12$, with solar
relative abundances.} \label{fig:narrow2phot}
\end{figure}

The velocity width of {\ion{O}{6}} ($b=9\pm1$\,\kms)\footnote{This
$b$-value was obtained by taking the second moment of the apparent
optical depth distribution over the velocity range $+30 \leq
v$[\kms]$ \leq +60$.} implies a maximum temperature of $\log
T$(K)$\leq4.9$. At this temperature collisional processes do not
produce appreciable amounts of {\ion{O}{6}}, so this high-ionization
gas likely arises from photoionization. The central engine is likely
to dominate the radiation field so we use the same ionizing spectrum
and range of ionization parameters used in the previous two
sections. The location of the gas is not clear since there is no
additional information (such as a velocity coincidence with a
line-emitting region, or a density diagnostic). As a result there is
fundamental uncertainty in the metallicity of the gas and, by
consequence, the total hydrogen column density. [Since \ion{H}{1}\
is not detected down to $N$(\ion{H}{1})$<1.8 \times
10^{13}$\,cm$^{-2}$ (see Table~\ref{tab:narrow}), we do not have a
means of inferring the metallicity. We can constrain the metallicity
only with a model for the ionization, and we return to this below.]
To place ionization constraints on the gas, we ran a grid of
\cloudy\ models over a range of ionization parameters ($-1 \leq \log
U \leq 1$) using the same ionizing spectrum and metallicity
([O/H]$=0.12$) as in the previous section for the \ion{O}{3}\ phase.
For each species, we generated a curve of total hydrogen column
density (as a function of ionization parameter) which reproduces the
observed species column density (or limit). The most constraining
curves (those for \ion{H}{1}, \ion{O}{5}, \ion{O}{6}, and
\ion{Ne}{8}) are shown in Figure~\ref{fig:narrow2phot}. The shapes
of the curves are determined by the ionization fraction of the
species, with the minimum in each curve occuring at the ionization
parameter that maximizes that species. The relative vertical
placements of the curves are determined by the relative abundances
of the elements (which are assumed to be solar in these models)
except for \ion{H}{1}. If the gas can be described by single
photoionized phase, then there must be a location (or region) on the
$N$(H)-$U$ plane that satisfies all column density measurements and
constraints. In the case of this component, such a region does
exist:
\begin{eqnarray}
(17.65) 17.75 \lesssim \log N(\mathrm{H}) \lesssim 18.1 (18.15) \\
(-0.45) -0.35 \lesssim \log U \lesssim -0.02 (0). \nonumber
\end{eqnarray}
The quoted range reflects the {$1\sigma$} confidence uncertainty
resulting from the error in $N$(\ion{O}{6}). The parenthetical
numbers indicate the $3\sigma$\ confidence range. The lower limit on
the ionization parameter is established by the comparison of
$N$(\ion{O}{6}) and the limit on $N$(\ion{O}{5}). This is
independent of assumed metallicity or relative abundance. The upper
limit arises from comparing the \ion{O}{6}\ and \ion{Ne}{8} column
densities which relies on the assumed relative Ne/O abundance
\citep[taken to be Ne/O$=0.151$,][and references
therein]{asplund04}. The limits on the ionization parameter are
probably not sensitive to the uncertainty in the metallicity since
the absorber is optically thin (in spite of the large total hydrogen
column density) to the ionizing radiation.

If the derived range of ionization parameter for the gas is
accurate, then we can place a lower limit to the metallicity of the
gas given the \ion{O}{6} column density and the \ion{H}{1} column
density non-detection limit via:
\begin{eqnarray}
[O/H] & = & \log \left [ { {N(\mathrm{O})} \over {N(\mathrm{H})} }
\right ] - \log \left [ { {N(\mathrm{O})} \over {N(\mathrm{H})} }
\right ]_\odot \\
& = & \log \left [ \left ( { {N(\mathrm{O\,VI})} \over
{N(\mathrm{H\,I})} } \right ) \left ( { {f_{\mathrm{O\,VI}}^{-1}(U)}
\over {f_{\mathrm{H\,I}}^{-1}(U)} } \right ) \right ] + 3.34,
\nonumber
\end{eqnarray}
where $f(U)$\ is the ionization fraction and we have used the
\citet{asplund04} value for the solar oxygen abundance. The minumum
value for ${
{f_{\mathrm{O\,VI}}^{-1}(U)}/{f_{\mathrm{H\,I}}^{-1}(U)} }$\ arises
at the largest allowed ionization parameter. For $\log U \sim
-0.02$, this fraction is $2.2 \times 10^{-5}$. Using the integrated
\ion{O}{6} column density and \ion{H}{1} column density limit from
Table~\ref{tab:narrow}, we find a limiting metallicity of [O/H]$\geq
-0.62$.

For a given absorber density, the $N$(H) axis can also be
interpreted as the thickness of the slab and $U$\ as the distance
between the absorber and quasar central engine. Using the limiting
metallicity, the total hydrogen column density could be as large as
$N$(H)$\sim 5 \times 10^{18}$\,cm$^{-2}$.

It is noteworthy that over the range of ionization parameters
allowed by the column density constraints, the dominant stage of
oxygen is \ion{O}{7}. For this component, our models predict
\ion{O}{7}\ column densities in the range $\log
N$(\ion{O}{7})$=14.4-14.8$.

\subsection{Broad Components}
\label{sec:broad}

\begin{figure*}
\epsscale{0.85}
\rotatebox{-90}{\plotone{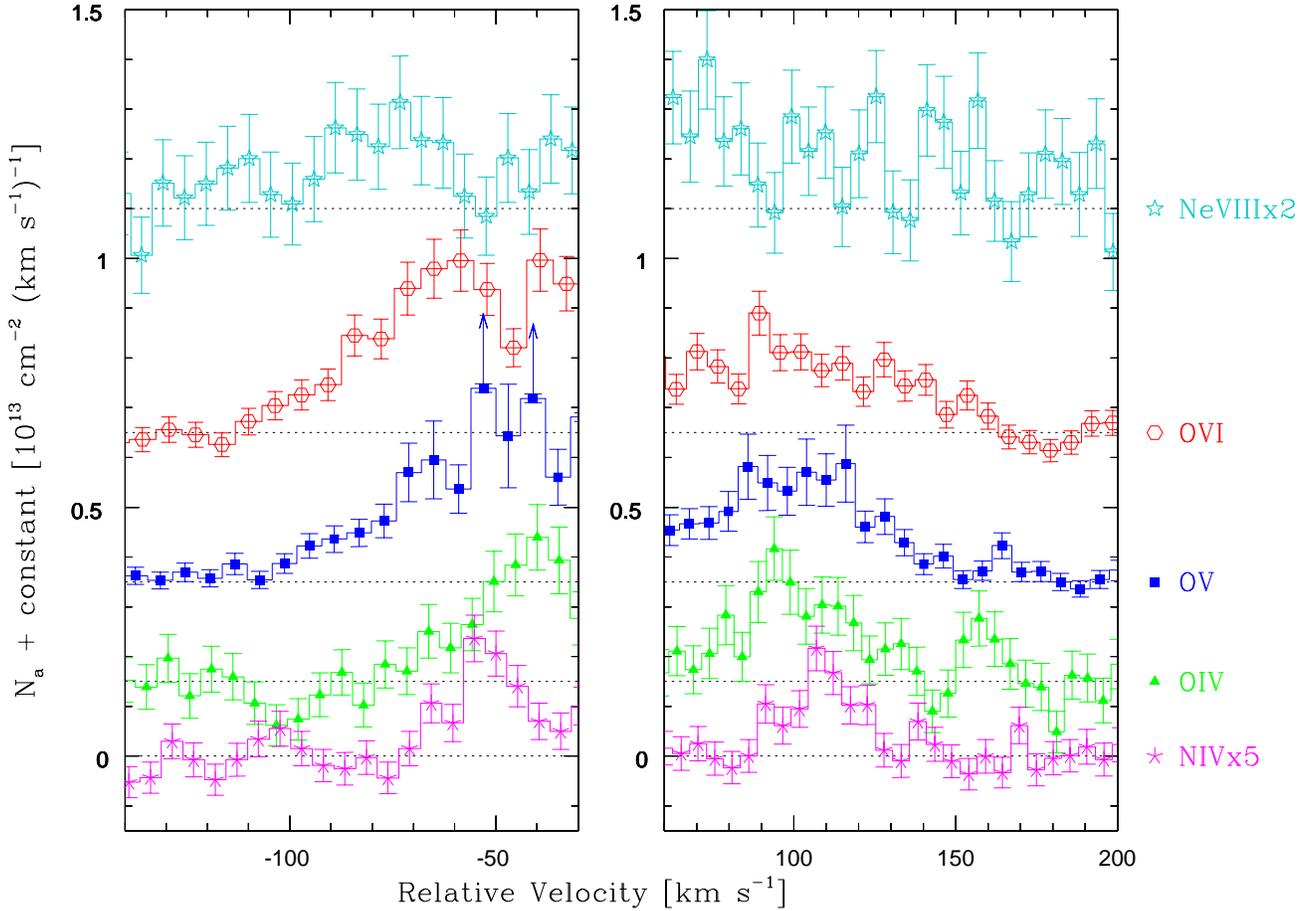}}
\protect\caption[Kinematics of Broad Components]{In the above
panels, we overlay the velocity-aligned apparent column density
profiles of the \ion{N}{4}, \ion{O}{4-VI}, and \ion{Ne}{8}\ ions on
blue and red wings (left and right panels, respectively) of the
associated absorber. For purposes of clarity, the profiles are
offset vertically, with the ``zero-level'' for each profile shown as
a horizontal dotted line. The \ion{O}{6}\ profiles are shown with a
sampling of 7\,\kms\ per bin, similar to the sampling the FUSE data.
In addition, the \ion{N}{4} and \ion{Ne}{8} profiles are shown with
a multiplicative factor to offset the abundance differences relative
to oxygen.} \label{fig:broad}
\end{figure*}

On the blue and red wings of the associated absorption-line system,
we detect broad components in the higher-ionization species:
\ion{N}{4}, \ion{O}{4-VI}, and \ion{Ne}{8}. Neither \ion{S}{4}, nor
any species with an ionization stage smaller than four is detected.
In Figure~\ref{fig:broad}, we show overlays of the \ion{O}{4-VI},
\ion{N}{4}, and \ion{Ne}{8} apparent column density profiles in the
velocity ranges $-140 \leq v \leq -20$\,\kms\ and $+60 \leq v \leq
+200$\,\kms. In Table~\ref{tab:broad}, we present measurements
(equivalent widths and integrated apparent column densities) of the
ions covered by our data over those two velocity ranges.

\input{tab3.tex}

Obtaining velocity centroids and widths for these components in a
non-parametric way using the apparent optical depth method is a
difficult task for two reasons. Both of these quantities involve
moments of the apparent optical depth and can be greatly affected by
the noise when the absorption is relatively weak (e.g., with
equivalent width less than five times the error), and by the choice
of integration range (which itself is affected by the assumed
structure of the absorption profile). For \ion{O}{5} $\lambda$629.73
and \ion{O}{6} $\lambda$1031.926, which are the strongest unblended
lines in these components, we find $b\sim23-32$\,\kms\ for the blue
wing and $b\sim29-40$\,\kms\ for the red wing using the integration
ranges quoted above and in Table~\ref{tab:broad}. These ranges
include an assessment of the uncertainty due to the choice of
integration range by adding and removing 20\,\kms\ to the range.

On the red wing, the apparent column density profile of
\ion{O}{4-VI} appear to match, at least to within the errors. There
is a possible excess of column density in the \ion{O}{5}\ profile at
$v = +110$\,\kms\ over \ion{O}{6}, and an excess of \ion{O}{6} at $v
= +140$\,\kms\ over \ion{O}{4}. The \ion{O}{4}\ profile is
consistent with zero column density at $v = +140$\,\kms. It is not
clear if this indicates an actual termination of the broad component
(at a velocity smaller either the \ion{O}{5}\ or \ion{O}{6}) with an
additional component at higher velocities, or whether both are
actually part of the same component. The similarity of the
\ion{O}{5} and the \ion{O}{6}\ profiles (modulo the aforementioned
excess) indicates that both are produced in the same gas. It is
difficult to place interesting ionization constraints on this phase
given the lack of additional information.

On the blue wing, there are clear differences in the kinematics of
the \ion{O}{4}\ profiles compared to \ion{O}{5}\ and \ion{O}{6},
which appear to rule out the production of all three species in the
same phase. The \ion{O}{4}\ profile appears to terminate at $v =
-80$\,\kms, with no detectable column density blueward of that
velocity. On the other hand, \ion{O}{5}\ and \ion{O}{6}\ are
detected at those velocities and appear to trace each other
perfectly at $v \leq -90$, with the termination of the profile near
$v = -110$\,\kms. In the range $-90 \leq v \leq -60$\,\kms, there is
a clear systematic excess of \ion{O}{6}\ over \ion{O}{5}. We note,
however, that the \ion{O}{5} $\lambda$629.730 line becomes optically
thick ($\tau = 1$) at an apparent column density of $1.2 \times
10^{12}$\,cm$^{-2}$\,(\kms)$^{-1}$, whereas the \ion{O}{6}
$\lambda$1031.927 line becomes optically thick at $2.8 \times
10^{12}$\,cm$^{-2}$\,(\kms)$^{-1}$. So the differences in the
profiles may be due to the effects of unresolved saturation.

We assume from the kinematic similarity of the \ion{O}{5} and
\ion{O}{6} apparent column density profiles that the two species
arise in the same gas. With only two ions with broad kinematics,
however, it is difficult to motivate a single ionization mechanism.
Thus, we consider both collisional ionization equilibrium (CIE) and
photoionization equilibrium (PIE) processes. In either case, the
ionization level (determined by temperature in CIE, or by ionization
parameter in PIE) and total oxygen column density can be gauged from
the \ion{O}{6}/\ion{O}{5} column density ratio. Inspection of the
integrated column densities in Table~\ref{tab:broad} reveals that
the \ion{O}{6}/\ion{O}{5} ratio is slightly higher on the blue wing
than the red wing, implying slightly higher ionization, although in
both cases the ratio is nearly unity. In CIE, this is easily
achieved at a temperature of $\log T$[K]$\sim5.5$. Photoionization
by the quasar can also produce such a ratio with an ionization
parameter of $\log U \sim -0.9$. We note that, in either case,
thermal broadening produces a line width that is smaller than
observed [$b_\mathrm{thermal}$(O)$\lesssim18$\,\kms\ vs. $b \sim
23-40$\,\kms], so there are likely other contributions to the
kinematics.

\begin{figure*}
\epsscale{0.75}
\plotone{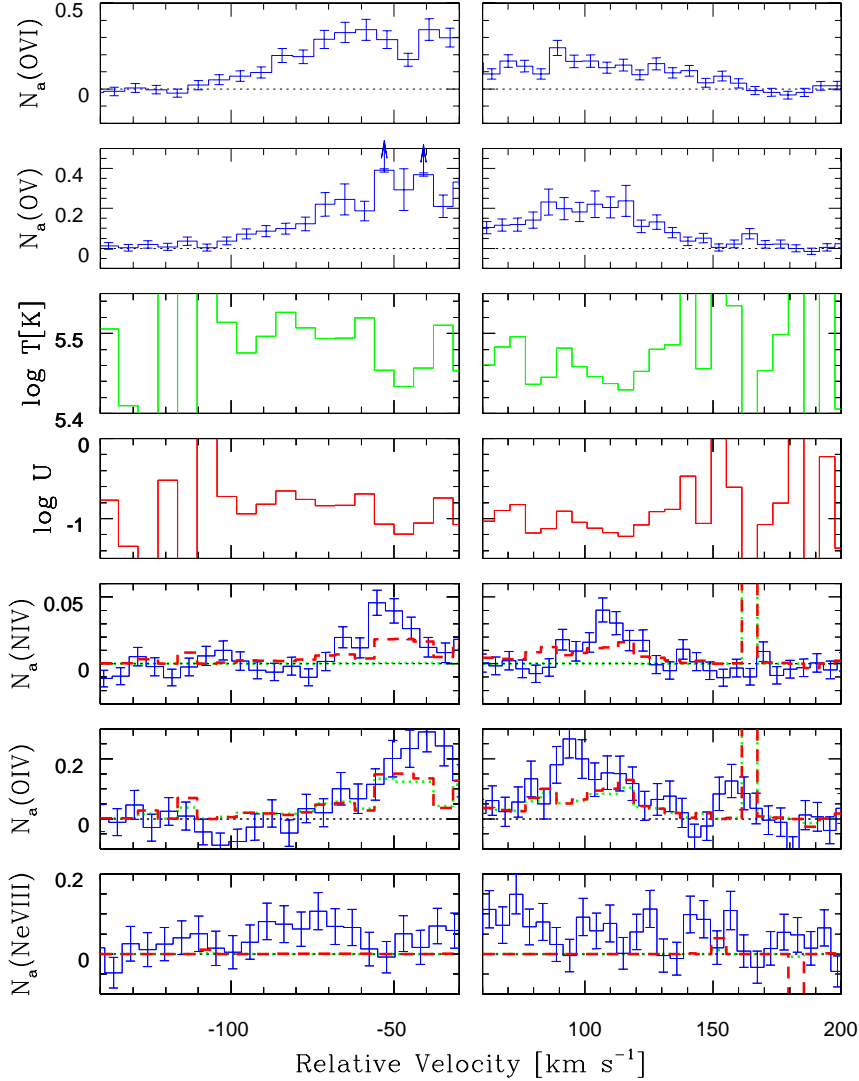}
\protect\caption[Ionization of Broad Components]{In the above
panels, we show the predictions from collisional ionization
equilibrium (CIE) and photoionization equilibrium (PIE) models on
the apparent column density profiles of \ion{N}{4}, \ion{O}{4}, and
\ion{Ne}{8}, where the ionization level is tuned to match the
\ion{O}{5-VI} profiles. On the left we show the analysis of the
broad component on the blue wing of the absorption-line system. On
the right, we show the same for the component on the red wing. The
top two panels on each side show the \ion{O}{6}, and \ion{O}{5}
apparent column density profiles. (Units on all apparent column
density profiles are $10^{13}$\,cm$^{-2}$\,(\kms)$^{-1}$.) The next
two panels show to the optimal temperature (for CIE) and ionization
parameter (for PIE) as a function of velocity that reproduces the
\ion{O}{6}/\ion{O}{5} column density ratio. In the bottom three
panels, we show the observed apparent column density profiles of
\ion{N}{4}, \ion{O}{4}, and \ion{Ne}{8} (solid histogram with error
bars), and the predicted profiles from CIE (dotted histogram) and
PIE (dashed histogram) models and assuming solar relative abundances
(N/O=0.132, Ne/O=0.151).} \label{fig:broad2}
\end{figure*}

In Figure~\ref{fig:broad2}, we compare the predicted \ion{N}{4},
\ion{O}{4}, and \ion{Ne}{8} apparent column density profiles from
CIE and PIE models with the observed profiles. These profiles were
generated by first assessing the temperature profile (for CIE) or
ionization parameter profile (for PIE) that reproduces the
\ion{O}{6}/\ion{O}{5} ratio at each velocity bin. The apparent
column density profiles were computed from the ionization fractions
implied by the temperature/ionization parameter profiles and assumed
solar relative abundances from \citet[][N/O=0.132,
Ne/O=0.151]{asplund04}:
\begin{equation}
N(\mathrm{X,i}) = N(\mathrm{O}, 5) \left ( f(\mathrm{X,i}) \over
f(\mathrm{O}, 5) \right ) \left ( \mathrm{X} \over \mathrm{O} \right
),
\end{equation}
where $N$(X, i) is the column density (or apparent column density
per unit velocity) of the ith ionization stage of element X, $f$(X,
i) is the ionization fraction of that ion, and (X/O) is the
abundance of X relative to oxygen. (This is a more general recasting
of equation 8.) All quantities are taken to be functions of
velocity. As mentioned above, the ionization fractions are
determined by temperature in CIE models, or by ionization parameter
in PIE models. The temperature profile was computed using the
collisional ionization equilibrium tables from \citet{sd93}. The
ionization parameter profile was computed using a grid of \cloudy\
models assuming the quasar radiation field. In both ionization
scenarios, detectable (and similar) amounts of \ion{O}{4} are
produced, though not enough to fully explain the observed profile.
Under the assumption of solar relative abundances, the CIE model
does not produce \ion{N}{4} (at the requisite temperature to satisfy
the \ion{O}{6}/\ion{O}{5} ratio) while the PIE model produces small
amounts. At least some of the \ion{O}{4} and \ion{N}{4} must reside
in a separate phase of gas.

It is noteworthy that neither model of the \ion{O}{5-VI}-bearing gas
is able to produce any \ion{Ne}{8} (under the assumption of solar
relative abundances) as shown in the bottom panel of
Figure~\ref{fig:broad2}. This may be an indication (in addition to
the lack of kinematic similarities) that \ion{Ne}{8} is produced in
a completely separate, very highly ionized phase of gas. Such a
phase would probably have a temperature of $\log T$[K]$\sim 5.85$,
if collisionally ionized, or $\log U \sim +0.4$, if photoionized. At
these ionization levels, all the oxygen would be in the form of
\ion{O}{7}. Assuming the solar relative abundance, we predict column
densities in the range $\log N$(\ion{O}{7})[cm$^{-2}$]$=15-15.5$ for
each of the two wings. If the metallicity of this very highly
ionized phase is solar, then the total hydrogen column density for
each wing is $\log N$(H)[cm$^{-2}$]$=18.3-18.8$. However, the origin
of this gas, and therefore its metallicity, is unclear. It could
arise near the quasar, or far from it in the halo of the host
galaxy.

\input{tab4.tex}

An alternative to a separate phase of gas that could potentially
explain the observed column densities/profiles of this wide range is
non-equilibrium processes. For example, from an apparent correlation
between $N$(\ion{O}{6}) and $b$(\ion{O}{6}) \citet{heck02} propose
that \ion{O}{6} arises in post-shocked, radiatively cooling gas.
Using cooling functions from \citet{sd93}, they predict column
densities of other ions that exist in both hotter gas (like that
producing \ion{Ne}{8}) that would exist prior to \ion{O}{6}, and
cooler gas (like that producing \ion{O}{5}) that exists after. In
Table~\ref{tab:cine}, we report the observed \ion{Ne}{8}/\ion{O}{6},
and \ion{O}{5}/\ion{O}{6} column density ratio for the broad
components on the blue and red wings. In addition, we report the
predicted column density ratios from \citet{heck02} for gas behind a
600\,\kms\ shock cooling from $T=10^6$\,K and a post-shock velocity
of 100\,\kms\ in two extremes, isobaric and isochoric cooling. From
the table, only the \ion{Ne}{8}/\ion{O}{6} column density ratio on
the red wing could be explained through a radiative cooling scenario
somewhere between these two extremes. However, in both components,
the observed \ion{O}{5}/\ion{O}{6} ratio is too large for either
extreme; there is too much \ion{O}{5} (relative to \ion{O}{6}) in
the observed profile.

Another alternative non-equilibrium process is photoionization by a
local radiation field produced by a fast, highly radiative, magnetic
pressure supported shock. \citet{ds96} present such a scenario and
provide tables of predicted column densities of a multitude of ions
for a range of shock velocities (200-500\,\kms) and magnetic
parameters, $B/n^{1/2}$\ (0-4 $\mu$G cm$^{3/2}$). The range of
\ion{Ne}{8}/\ion{O}{6} and \ion{O}{5}/\ion{O}{6} ratios predicted by
these models is also listed in Table~\ref{tab:cine}. In this case,
the \ion{Ne}{8}/\ion{O}{6} ratio of both components could be
explained under this scenario. However, as with radiative cooling
scenario, the predicted \ion{O}{5} column relative to \ion{O}{6} is
much smaller (by about a factor of two) than observed. Other
non-equilibrium scenarios include turbulent mixing of entrained gas
\citep*[e.g.,][]{ssb93}, and magnetic conduction through the
interface between gases \citep*[e.g., ][]{bbf90}. However, these
papers make predictions on the column density of more commonly
detected ions, such as \ion{C}{4}, \ion{N}{5}, and \ion{Si}{4} which
are not covered by our data. Thus we cannot address the
applicability/validity of these scenarios to the gas in the broad
components.

\section{Discussion}
\label{sec:discussion}

We have analyzed the detailed ionization and geometric constraints
and physical conditions provided by the wide range of ionization
species detected in the absorption line system associated with the
quasar HE\,0226-4110. The rest-frame wavelength coverage of our FUSE
and HST/STIS data, 610--1150\,\AA, provide coverage of six
ionization stages of oxygen. Four stages are detected
(\ion{O}{3-VI}) and these show a remarkable ionization-dependent
complexity in the gas kinematics. Comparatively low-ionization
species like \ion{H}{1}, \ion{C}{3}, and \ion{O}{3} are only
detected in a single narrow component. Two broad components flank
this component in velocity and appear mostly in high-ionization
species. Intermediate-ionization stages, like \ion{N}{4} and
\ion{O}{4}, are featured in both a narrow component (although with
in a separate phase that is offset in velocity) and the two flanking
broad components. An additional narrow component exists in this
associated system that is only apparently detected in \ion{O}{6}. In
the following sections we discuss possible origins, locations, and
structural geometries for these components. [As has been our
convention through-out this paper, we quote velocities relative to
$\zabs=0.4925$.]

\subsection{Low and Moderate-ionization gas at $v=-8$\,\kms\ (in the $\zabs=0.4925$\ rest-frame)}

The narrow low and moderate-ionization component detected in
\ion{H}{1}, \ion{C}{3}, and \ion{O}{3}\ is the most well-constrained
component in this absorption-line system. All three species appear
to arise in the same phase of gas, as motivated by the kinematics of
profiles. Using measurements of the \ion{H}{1}/\ion{O}{3} and
\ion{C}{3}/\ion{O}{3} ratios (and limits on the
\ion{O}{4}/\ion{O}{3} and \ion{N}{4}/\ion{O}{3} ratios), we
determine the optimal metallicity, ionization parameter and total
hydrogen column density of the \ion{O}{3}-bearing gas (see
equation~\ref{eq:narrow1phot}). We now focus on the implications of
these results on the location and geometry of the gas.

The ionization parameter is related to the spectral energy
distribution of the quasar ($L_\nu$), the distance between the
absorber and the quasar central engine ($r$), and the density of the
absorber ($n$) via:
\begin{equation}
U = {1 \over {4 \pi r^2 n c}} \int_{\nu_{\mathrm{LL}}}^{\infty}
{{L_\nu} \over {h \nu}} d\nu,
\end{equation}
where $\nu_{\mathrm{LL}}$\ is the frequency of the Lyman limit. For
the spectral energy distribution specified in the photoionization
models, this relationship reduces to:
\begin{eqnarray}
\log U & = & -1.43 - \log n_{3.5} - 2 \log r_{21.5} \label{eq:urreal} \\
n_{3.5} & = & n/(10^{3.5} \mathrm{cm}^{-3}) \nonumber\\
r_{21.5} & = & r/(10^{21.5} \mathrm{cm}) = r/(1 \mathrm{kpc}).
\nonumber
\end{eqnarray}
The reference values for the density and distance are taken from
current limits for intrinsic narrow absorption lines based on
variability analyses
\citep[e.g.,][]{wise04,narayanan04,gan01c,ham97b}.

Likewise, the thickness scale of the absorbing gas is given by the
ratio between the total hydrogen column density and the density of
the gas. Combining equations~\ref{eq:narrow1phot} and
\ref{eq:urreal}, we find the following relationships for the
location ($r$) and thickness ($\Delta r$) of the absorbing gas:
\begin{eqnarray}
r = 1.5 n_{3.5}^{-1/2}~\mathrm{kpc} \label{eq:oiiigoem}\\
\Delta r = 7.3 n_{3.5}^{-1}~\mathrm{AU}. \nonumber
\end{eqnarray}

Before considering possible locations for the gas and related
implications, we make one observation regarding the spatial extent
(i.e., perpendicular to the line of sight) of the absorbing
structure. Given that the absorber fully occults the UV continuum as
evidenced by the zero flux levels of the \ion{C}{3}
$\lambda977.020$\ and \ion{H}{1} Lyman $\beta$\ lines, we can place
a lower limit on extent of the absorber on the plane of the sky.
Using our fit to the optical spectrum (in particular, our
characterization of the H$\beta$\ emission line and the underlying
power law continuum), we estimate that the black hole powering the
quasar has a mass of $10^{8.7-9.6}$\,M$_\odot$. The range comes from
using the scaling relations of \citet{kaspi00} and
\citet{peterson04}. Black holes with masses in that range have a
gravitational radius of $r_{\mathrm{g}}=$4.6--38\,AU. The UV
continuum is typically emitted in the region between
6$r_{\mathrm{g}}$\ and 60$r_{\mathrm{g}}$ {\citep[e.g., ][]{mur95}},
with the bounds arising from the radius of last stable orbit (for a
non-rotating black hole) and the radius at which the accretion-disk
becomes optically thick. Thus the absolute minimum spatial extent of
the cloud is $w\approx 600$\,AU ($2 \times 60 r_{\mathrm{g}}$).
Combining this with equation~\ref{eq:oiiigoem}  The aspect ratio of
the \ion{O}{3}-bearing gas (width/thickness) is then:
\begin{eqnarray}
w/\Delta r \gtrsim & 82 n_{3.5} \label{eq:oiiiasp} \\
w/\Delta r \gtrsim & 82 \left ( {r \over {1.5\,\mathrm{kpc}}} \right
)^{-2} \nonumber
\end{eqnarray}
In the second expression, we have used the relationship between the
source-absorber distance, and the the absorber density.

From a consideration of the statistical frequency of high-ionization
(\ion{C}{4}, \ion{N}{4}, and \ion{O}{6}) associated absorbers in
low-redshift ($z \lesssim 1$) quasars as a function of quasar
redshift, \citet{gan01a} placed high-ionization AAL gas in the
shearing region between the outflowing from the accretion disk, and
a low-density, very highly ionized medium above the wind (see their
Figure~13). This scenario was physically motivated by the
hydrodynamic simulations of \citet{psk00} which show instabilities
in that region. For the \ion{O}{3}-bearing gas to reside in that
region (within a parsec of the UV continuum source), the gas must
have a density in excess of $7 \times 10^9$\,cm$^{-2}$, and an
aspect ratio larger than $1.8 \times 10^8$. While the density is
fairly reminiscent of the broad line regions of quasars, the aspect
ratio is unphysically large. Similarly, the placement of intrinsic
narrow absorption lines by \citet{elvis00} at the bottom of the
outflow would require even larger densities and aspect ratios. This
does not necessarily rule out either model, since those models only
apply to high ionization absorbers. Rather, this AAL, and in
particular the \ion{O}{3}-bearing gas, points to an incompleteness
in the models purporting to place gas that is related (at least
statistically) to quasars. In the following subsections, we
entertain two possible, alternative locations for the gas producing
the \ion{O}{3}-bearing gas, the narrow emission line region of the
quasar, and the halo of the quasar host galaxy.

\subsubsection{Narrow Emission Line Gas?}

In principle, combining information from [\ion{O}{3}] $\lambda$5007
emission and \ion{O}{3} $\lambda$832.927 absorption, we can
potentially constrain the density ($n$) of the absorbing gas.
Density information is important since it enables an assessment of
the geometry of the absorbing material courtesy of equation~9. In
some associated absorbers, density constraints (e.g., from excited
fine-structure lines) combined with photoionization models have
indicated large distances between the associated absorber and the
QSO central engine \citep[e.g.,
][]{morris86,tls96,ham01,ehc03,gabel04}.  In other cases, similar
analyses place the AALs close to the central engine \citep[e.g.,
][]{sp00}. The density estimates on these rely on excited states of
comparatively low-ionization species, namely \ion{C}{2}$^*$,
\ion{Si}{2}$^*$, \ion{Fe}{2}$^*$, and \ion{C}{3}$^*$\ that are not
widely detected in AALs, so it is not clear how broadly applicable
the results of those studies are in understanding the distribution
of location of AAL gas. The potential association of \ion{O}{3}
absorption in the narrow-emission line region, then, has paramount
importance to this issue, since strong narrow emission lines are far
more common in AGN.

In trying to associate this absorption-line gas with the narrow
emission line gas observed in [\ion{O}{3}] $\lambda$5007, it is
important to consider the geometry and kinematics of the emitting
region compared to the pencil-beam probe producing the absorption.
We note a few basic facts (in addition to those mentioned above
regarding physical conditions and geometry) that a narrow emission
line region model incorporating the \ion{O}{3}-absorbing gas should
address: (1) a match in velocity ($\zem$([\ion{O}{3}])$=0.4928 \pm
0.0001$, $\zabs$(\ion{O}{3})$= 0.49246 \pm 0.00006$, $\Delta v=68
\pm 23$\,\kms; see Table~\ref{tab:emission} and \S\ref{sec:aal})
between the absorption and emission line centroid; (2) a factor of
32 difference in the velocity widths (FWHM$\sim$470\,\kms\ for
emission, FWHM$\sim$15\,\kms\ for absorption); and (3) an
[\ion{O}{3}] $\lambda$5007 luminosity of $3.2 \times 10^{42}$\,erg
s$^{-1}$.

We first consider whether gas with $\log U = -2.29$\ and [O/H] $=$
+0.12 (as implied by the analysis of the absorption lines) can
reproduce the observed luminosity. To do so, we ran another grid of
\cloudy\ models over density, placing the gas at a distance that
reproduces an ionization parameter of $\log U = -2.29$. At each grid
point, we compute the [\ion{O}{3}] $\lambda$5007 line luminosity and
compare that to the observed line luminosity for emission-line
component 6 from our fit to the [\ion{O}{3}] $\lambda$5007 line in
\S\ref{sec:quasar} (see Table~\ref{tab:emission}), since this is the
component whose redshift matches that of the \ion{O}{3}
$\lambda$832.927 absorption. In these models, the gas is assumed to
exist in the thin spherical shell that surrounds the point-like
central engine. The absorption arises from the pencil-beam probe
through the shell toward the observer, while the emission arises
(isotropically) from the integration over the entire shell.

\begin{figure}
\epsscale{1.1}
\rotatebox{-90}{\plotone{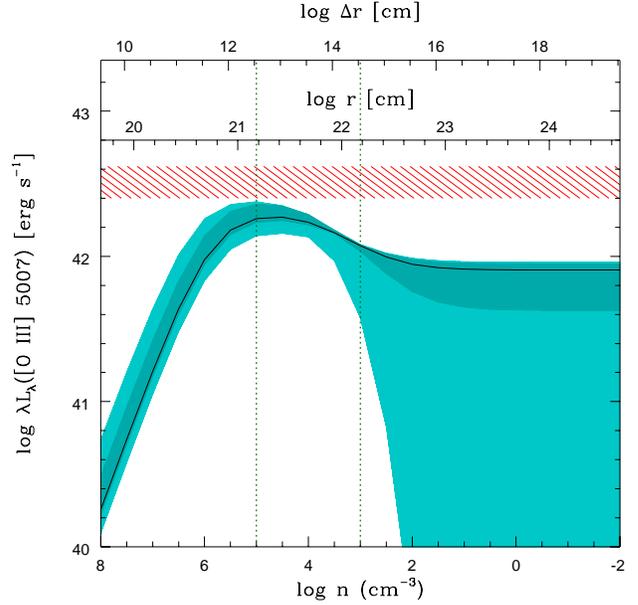}}
\protect\caption[Phtoionization of $v\sim0$\,\kms\ component]{In the
above figure, we plot the predicted [\ion{O}{3}] $\lambda$5007 line
luminosity as a function of gas density. The solid curve shows the
grid of \cloudy\ models using $\log U = -2.29$, and [O/H]=+0.12. The
shaded regions around the solid curve represent the set of models
allowed by the $1\sigma$\ and $2\sigma$\ uncertainties in the
ionization parameter and metallicity from the absorption-line
diagnostics (see Figures~\ref{fig:narrow1a}-\ref{fig:narrow1phot1}).
The hatched region at $\log \lambda L_\lambda$([\ion{O}{3}]
$\lambda$5007)[erg~s$^{-1}$]=42.4-42.62 marks the $1\sigma$
confidence luminosity of the EM6 emission line component which we
associate with this gas. The dashed vertical lines at $\log
n$[cm$^{-3}$]=3 and 5 indicate the typical range of densities
expected for the narrow emission line region gas using emission-line
diagnostics. The top two axes indicate the inferred absorber
thickness ($\Delta r$, assuming a total hydrogen column density
$\log N$(H)[cm$^{-2}$]$=17.54$) and distance from the accretion disk
($r$). These axes are implicitly dependent on the ionization
parameter, and therefore only apply to the solid curve.}
\label{fig:narrow1phot2}
\end{figure}

The results of the grid of \cloudy\ models are shown in
Figure~\ref{fig:narrow1phot2}. On the top axes of the plot, we show
the implied distance, $\log r[\mathrm{cm}] = 21.5 + 0.5 (4.36 - \log
n[\mathrm{cm}^{-3}])$, and thickness, $\log \Delta r[\mathrm{cm}] =
17.54 - \log n[\mathrm{cm}^{-3}]$. The model curve shows a power-law
decrease of the [\ion{O}{3}] $\lambda$5007 line luminosity with
increasing density starting at $\log n[\mathrm{cm}^{-3}] \sim 5$. At
smaller densities, the curve falls off and flattens to a constant
line luminosity whose value depends on the ionization parameter and
metallicity. The shaded regions around the plotted curved represent
families of models that lie in the $1\sigma$\ and $2\sigma$\
confidence regions shown in Figure~\ref{fig:narrow1phot1}. Models
with higher ionization parameters and/or metallicities decrease the
constant luminosity to which the curve levels out. The hatched
region across the top of the figure is vertically placed at the
observed luminosity of the EM6 component, with the thickness of the
region representing the $1\sigma$\ confidence interval. The vertical
dotted lines are placed to indicate the range of densities, {$\log
n[\mathrm{cm}^{-3}] = 3 - 5$, commonly associated with narrow
emission line region gas. The plot shows that there is no density at
which the model predicts sufficient line luminosity to match the
observation, although at the $\sim1.5\sigma$\ level, there is
reasonable agreement around $n \sim 10^{5}$\,cm$^{-3}$.

In principle, the discrepancy could be further explained by the
fragmentation of the shell into several smaller clouds within the
narrow emission line region (with only one cloud producing the
absorption) or with the inclusion of other important physical
processes (such as shocks). Such an explanation may be required in
order to reconcile the model with the observed velocity width of the
EM6 component ($\sim$470\,\kms). A static thin shell with an
internal velocity dispersion of $\sim$15\,\kms\ (as indicated by the
pencil-beam absorption-line probe) does not yield an overall
velocity width of 470\,\kms\ from the emission-line gas. [In order
to explain the 68\,\kms\ blueshift of the absorption line relative
to the emission line, one might invoke an expanding thin shell, but
even such a shell would not produce the observed emission line
velocity width.] With a fragmented shell, a possible solution is
that the bulk velocity dispersion of the resulting clouds (arising
perhaps from orbital motion about the central black hole) is several
hundred \kms, while the internal velocity dispersion of any single
cloud is as indicated by the absorption-line profiles. This scenario
has two advantages over a simple thin shell model: (1) a thin shell
may have stability problems, and (2) the added surface area from
fragmentation could yield the factor of two difference between the
predicted and observed [\ion{O}{3}] $\lambda$5007 luminosities.

However, there are other potential self-consistency problems with
the absorbing gas with a density of $n\sim10^5$\,cm$^{-3}$. The
implied distance between the absorber and the accretion disk is $r
\approx 0.5 n_{5}^{-1/2}$\,kpc. While this is reasonable for a
location within the narrow emission line region
{\citep[e.g.][]{schmitt03,bennert02}}, the implied cloud thickness
is extremely small, $\Delta r \approx 0.24 n_{5}^{-1}$\,AU, implying
a minimum aspect ratio (see equation~\ref{eq:oiiiasp}) of $\approx
1200 n_{5}$\ for the cloud. For the model to provide a
self-consistent representation of data, the gas probed by the
absorption must be distributed in in the form of a sheet or a closed
spherical shell (or sets of sheets/shells), rather than a discrete
almost-spherical cloud. Possibly, such a deformation might result
from the hydrodynamics of gas orbiting a central supermassive black
hole, although such a deformed cloud would probably fragment
further. Another potential problem with this scenario is that only
one cloud is seen in absorption. While it is possible that only one
cloud would be intercepted by the sightline in this patchy cloud
scenario, it is improbable. Only future observations to look for
variability in the \ion{O}{3}-absorption profile and/or higher
resolution observations (spatial and spectral) of the [\ion{O}{3}]
$\lambda$5007 can address the patchiness of the narrow emission line
region and the number of clouds intercepted by the absorption-line
probe.

A number of additional observational tests are required to fully
explore the implications of this result. Additional data covering
the \ion{O}{3} $\lambda$832.927 line in the associated absorption of
other quasars is required to ascertain the frequency in which gas in
the narrow emission line region is observed in absorption and the
fraction of associated absorbers that originate in this region.
Since this associated absorber clearly has many components with
different gas phases, it is likely that several different regions
contribute to the observed profiles. Spectra with high-resolution
($\lambda/\Delta \lambda \geq 10^4$) are required to disentangle the
relative importance of the different quasar emitting regions to the
observed frequency of associated absorbers. For this and other
quasars whose spectra feature associated absorption in \ion{O}{3}
$\lambda$832.927), higher-resolution spectroscopy of the
[\ion{O}{3}] $\lambda$5007 line would be useful in directly, and
more precisely, comparing the kinematics of the extended narrow
emission-line region with the pencil-beam probe provided by the
\ion{O}{3} absorption. The density diagnostics available from the
analysis of emission lines would greatly assist in disentangling the
geometry of intrinsic absorbers.

\subsubsection{Quasar Host Galaxy Halo Gas?}

Another potential location for the \ion{O}{3}-absorbing gas is the
halo of the quasar host galaxy. If the gas were located, for
example, 10\,kpc away from the central engine, then (by
equation~\ref{eq:oiiigoem}) the density of the gas would be
$\sim$230\,cm$^{-3}$, and the thickness of the gas would be
$\sim$100\,AU. The resulting aspect ratio (by
equation~\ref{eq:oiiiasp}) would be $>6$, so such a structure would
still be rather non-spherical. Are there local analogues of such a
structure? High-velocity clouds such as Complex C or the Magellanic
Stream may have such aspect ratios, but do not have such large
densities
{\citep[e.g.,][]{sembach01,wakker01,tripp03,fox04,fox05,wakker04a}}.
If the gas were $\gtrsim$13.5\,kpc away from the central engine,
also within the halo of the quasar host galaxy, that would allow for
a spherical structure with an absorbing thickness scale $>$600\,AU.
At such a distance, the density is smaller than 37\,cm$^{-3}$.
However, clouds in the halo (or outer corona) of the Galaxy do not
have the high (super-solar) metallicity
{\citep[e.g.,][]{wakker01,richter01,tripp03,wakker04a} observed in
the \ion{O}{3}-absorbing gas.

While there are no local analogues of an absorbing structure that
exists in the halo with super-solar metallicites and a large
density, such an idea is not without precedent. The associated
absorber observed toward 3C\,191 \citep[][]{ham01} features very
similar properties - gas with a density of $\sim$300\,cm$^{-3}$,
residing 28\,kpc from the quasar central engine. However, that
absorber is detected in very low-ionization gas (\ion{Mg}{2},
\ion{C}{2}, \ion{Si}{2}) and excited state lines from \ion{C}{2}*
and \ion{Si}{2}*, which are not present in the \ion{O}{3}-absorbing
gas in this associated absorber. Moreover, the low-ionization lines
observed in the 3C\,191 associated absorber exhibit the signature of
partial coverage (dilution of the absorption line troughs unocculted
light) and the coverage fractions imply gas clouds with a spatial
extent smaller than 0.01\,pc. (The thickness of the 3C\,191
associated absorber is unknown as \citet{ham01}\ are only able to
place upper limits on the total hydrogen column density.) In
addition, the 3C\,191 associated absorber is apparently outflowing
(it is blueshifted by more than 400\,\kms\ relative to the broad
emission lines), whereas the \ion{O}{3}-absorbing gas in this
associated absorber is at a velocity fairly coincident with the
[\ion{O}{3}] $\lambda$5007 narrow emission line. It is possible that
\ion{O}{3}\ gas in this associated absorber is a slightly
higher-ionization analogue of that absorber. If so, then the
speculation by \citet{ham01}\ that the gas may be residue from a
nuclear starburst superwind may be applicable. It is unlikely that
this material is the result of an ejection from a BAL wind as the
gas appears to be at rest (or close to it) with respect to the
narrow emission region, which would imply a severe deceleration (by
several order of magnitude).

\subsection{Moderate and High-ionization gas at $v=$0 \& +50\,\kms\ (in the $\zabs=0.4925$\ rest-frame)}

Additional narrow features are detected in this associated
absorption-line system, but the physical conditions of the gas are
much less constrained than the \ion{O}{3}-bearing gas. Both of these
components have a higher ionization than the other narrow component.
The $v=0$\,\kms\ component is detected as a single narrow feature in
intermediate-ionization species such as \ion{N}{4}, \ion{O}{4}, and
\ion{S}{4}, while the $v=+50$\,\kms component is detected as a
narrow feature in \ion{O}{6} only. Absorption from other
intermediate- and high-ionization species - \ion{O}{5} and
\ion{Ne}{8} - is evident, but kinematics of these species do not
trace the narrow components. There is no detectable \ion{H}{1} that
is associated with either of these components, as all the observed
\ion{H}{1} appears over velocity range and has the kinematic
appearance of the \ion{O}{3}-bearing gas. Whether there is a
relationship between these components either to each other, or to
the \ion{O}{3}-bearing gas is unclear. The gas could lie close to
the quasar central engine, or far out in a companion galaxy.

The velocity widths of these two components are fairly narrow
[$b$($v=$0\,\kms)$\sim 15$\,\kms, $b$($v=$+50\,\kms)$\sim
9$\,\kms]\footnote{Both of these values were obtained from direction
integration of the second moment of the apparent optical depth
profile.} and, in the case of the $v=+50$\,\kms\ component, suggest
that collisional processes are not important in the ionization,
leaving radiative processes as the likely dominant source of
ionization. Even if the gas producing these components were far from
the central engine, perhaps in a dwarf companion to the quasar host
galaxy, photionization by the central engine is likely the dominant
source of ionizing photons. For the $v=+50$\,\kms\ narrow component,
we used the available constraints from \ion{O}{5} and \ion{Ne}{8} to
infer that the gas must have an ionization parameter in the range
$-0.35 \lesssim \log U \lesssim -0.02$. However, since there is no
information on the density of this component (e.g., via association
with a line-emitting region, or excited state transitions, or
time-variability), it is difficult to know where the gas is located
relative to quasar central engine [since the distance between the
accretion disk and the absorber is degenerate with the density ($U
\propto L/n r^2$)].

One possible interpretation is that these components arise in
structures similar to that of the other narrow component, but with
smaller densities (thus producing higher-ionization species). For
the narrow \ion{O}{6} component, if the gas lies within the narrow
emission line region of the quasar, then it has a density in the
range $10^{2.69} \lesssim n[\mathrm{cm}^{-3}] \lesssim 10^{3.02}$.
This is consistent with the range of densities typically found in
quasar narrow emission line regions. The gas producing the narrow
\ion{O}{6} absorption would not contribute significantly to the
formation of the narrow emission lines since the gas is too ionized.
If the total hydrogen column density of the narrow \ion{O}{6}
component is in the range $17.75 \lesssim \log
N(\mathrm{H})[\mathrm{cm}^{-2}] \lesssim 18.1$\ (as would be implied
if the metallicity were similar to the \ion{O}{3} component), then
the thickness of the gas sheet producing the narrow \ion{O}{6}
component lies in the range $35 \lesssim \Delta r[\mathrm{AU}]
\lesssim 170$, and an aspect ratio of at least 3.5. Using the
absolute maximum total hydrogen column density (as derived by the
non-detection of \ion{H}{1} and the limiting ionization fraction),
the aspect ratio cannot be smaller than 1.4 (if the gas lies with
the narrow emission line region).

Alternatively, if the narrow \ion{O}{6} gas is in the halo of the
quasar host galaxy, 10\,kpc away from the central engine, then it
would have a density in the range $0.5 \lesssim n[\mathrm{cm}^{-3}]
\lesssim 1.1$, which would be consistent with local versions of halo
gas clouds. While these are only two of many possible scenarios for
these higher-ionization components, they have the attraction of
readily explaining why all three narrow components arise within the
same absorption-line system very close in velocity (within
50\,\kms).

It is noteworthy that \citet{ber02} and \citet*{csk02} detect
\ion{O}{6} components at $\zabs \gtrsim 2$\ which are also narrow
enough to imply photoionization. In particular, \citet{ber02} detect
a component  at $\zabs=2.36385$ in the direction of Q 0329-385
($\zem=2.423$) with a similar $b$-value as the $v=+50$\,\kms\
component in this associated system. [The difference in redshift
corresponds to a velocity separation of 5225\,\kms, which is close
to the arbitrary cutoff for associated absorption-line systems.]
That component is detected in \ion{H}{1} and has a
$N$(\ion{O}{6})/$N$(\ion{H}{1}) ratio of $\approx 0.09$. Our narrow
\ion{O}{6} component is not detected in \ion{H}{1} and has a ratio
of $N$(\ion{O}{6})/$N$(\ion{H}{1})$\gtrsim 5$. In fact, in all of
the cases presented by both \citet{ber02} and \citet{csk02}, the
narrow, hence photoionized, \ion{O}{6} is accompanied by detection
of \ion{H}{1}, in marked contrast to the $v=+50$\,\kms\ component in
this associated absorption-line system. [\citet{csk02} note that in
all cases that they present, the \ion{H}{1} column density is larger
than $2 \times 10^{14}$\,cm$^{-2}$. None of their components appear
to have metallicities in excess of [C/H]$\approx-1.7$.]
\citet{ber02} report that their photoionized-\ion{O}{6} components
(with only one exception) have metallicities in range $-2.5 \leq
$[C/H]$ \leq -0.5$, with super-solar [O/C]. This is consistent with
our constraint on [O/H] ($\geq -0.62$) for the $v=+50$\,\kms\
component.

In many of the cases studied by \citet{ber02} (including the one
mentioned above), the narrow \ion{O}{6} components are accompanied
by detections of \ion{C}{4} (and occasionally \ion{N}{5}). For the
narrow \ion{O}{6} component in this absorption-line system, the
strong UV doublet from neither species is covered by our spectra. If
the photoionization models described in \S\ref{sec:narrow3} are an
accurate representation of the \ion{O}{6}-bearing gas, then we
predict column densities in the range $12.21 \lesssim \log
N$(\ion{C}{4})$ \lesssim 12.4$, and $12.74 \lesssim \log
N$(\ion{N}{5})$ \lesssim 12.82$\ (for [O/C]=[O/N]=0). For
comparison, the associated component from the \citet{ber02} study
mentioned above has $\log N$(\ion{C}{4})$=12.33$, and $\log
N$(\ion{N}{5})$=12.00$. The \ion{C}{4} column density is within the
range predicted for the $v=+50$\,\kms\ component (even if [O/C] were
slightly super-solar), but the \ion{N}{5}\ column density is smaller
than that of the $v=+50$\,\kms\ component (if N/O is solar). If the
N/O abundance in the $v=+50$\,\kms\ component is subsolar by 0.7
dex, then the two are components are very similar aside from the
differences in redshift and \ion{H}{1} column density. \citet{ber02}
find precisely such an abundance requirement for their
$\zabs=2.25146$\ component ([O/N]$\geq 0.7$). If this is a general
property, then the $v=+50$\,\kms\ component in this associated
absorption-line system is very consistent with the populations of
photoionized-\ion{O}{6} discussed by \cite{ber02} and may be higher
metallicity analogues to those studied by \cite{csk02}.


\subsection{Broad Components}

Intermediate and high-ionization species, namely \ion{N}{4},
\ion{O}{5-VI}, and \ion{Ne}{8}, show absorption in two relatively
broad components. In \ion{O}{6}, the absorption in these two
components constitutes roughly 30\% of the total integrated column
density. The kinematics of \ion{O}{5} and \ion{O}{6} appear to trace
each other, suggesting that the two arise in the same phase of gas.
Moreover, there is no evidence for trough-dilution by unocculted
quasar light, so it is unlikely that these components arise from the
quasar outflow (depending on the orientation of the accretion disk
and possible presence of a scattering medium).

Since many different physical process could be involved in the
production of this gas (photoionization by the quasar central
engine, collisional ionization from shocks, etc.), it is unclear
what ionization model to apply in order to infer physical
conditions. In \S\ref{sec:broad}, we explored various ionization
models, collisional ionization equilibrium (CIE), photionization
equilibrium (PIE), radiative cooling (RC) as describe by
\citet{heck02}, and shock ionization (SI) as described by
\citet{ds96}. CIE and PIE models that are tuned (through the
temperature or ionization parameter) to reproduce the observe
\ion{O}{5} and \ion{O}{6} column densities are unable to fully
explain the presence of \ion{N}{4}, \ion{O}{4}, and \ion{Ne}{8} in
the same phase of gas (Fig.~\ref{fig:broad2}). The RC and SI models
are unable to even simultaneously reproduce the \ion{O}{5} and
\ion{O}{6} column densities under the range of parameters considered
(see Table~\ref{tab:cine}).

Of the models considered, photoionization by the quasar is able to
explain the largest number of species simultaneously. However, such
a model requires additional phases to explain the presence of
\ion{Ne}{8}, and some of the \ion{N}{4} and \ion{O}{4} (and it is
not clear what ionization models are appropriate for such phases).
For the phase producing \ion{O}{5-VI}, if photoionization by the
quasar is the correct model, then \ion{O}{7}\ should be the dominant
ionization stage in these components. An X-ray spectrum (e.g, with
{\it Chandra} or XMM-{\it Newton}) to look for \ion{O}{7}\
absorption would help to elucidate the important physical processes
producing the broad components. Although the total column density of
\ion{O}{7}\ predicted by these components ($N \lesssim
10^{15}$\,cm$^{-2}$) would probably not be detectable, the higher
ionization gas producing the \ion{Ne}{8} may be detectable.
Information on the higher ionization stages of oxygen would compete
the picture of the ionization structure of gas in this
absorption-line system.

Like the narrow higher-ionization components, there is no density
information in the broad components. Thus, it is not possible to
transform the optimal ionization parameter into a distance.
Likewise, there is no metallicity information, so it is not possible
to robustly gauge the thickness of the absorbing regions. There are
several plausible locations for this gas (ablation off an obscuring
torus, a diffuse medium co-spatial with the narrow emission-line
region, satellite dwarf galaxies, etc.) and we cannot distinguish
between them.

\section{Summary of Results}
We have obtained high-resolution, rest-frame extreme ultraviolet
spectra of the quasar HE\,0226-4110. The spectra cover the
wavelength range 610--1150\,\AA\ in the rest-frame of the quasar. We
detect an associated absorption line system in a wide range of
ionization species, including several Lyman series lines,
\ion{Ne}{8}, and four adjacent stages of oxygen, \ion{O}{3-VI}.
Strong transitions from \ion{O}{1}\ and \ion{O}{2}\ are also covered
by our data, but are not detected. The high quality of these spectra
allow us to study the kinematic structure of the gas. A summary of
our results follows.

\begin{itemize}
\item[1.] The \ion{O}{3} $\lambda$832.927 line is detected in a
single narrow component at $v=-8$\,\kms\ in the $z=0.4925$\
reference frame. The kinematics of the absorber indicate that all
the detected \ion{H}{1}\ and \ion{C}{3}\ are associated with this
component, and we use the \ion{O}{3}\ profile to gauge the column
density ratios of the narrow emission line region gas. Using
\cloudy\ photoionization models to constrain physical parameters, we
find that this cloud has a metallicity of
[O/H]$=+0.12_{-0.03}^{+0.16}$, an ionization parameter $\log
U=-2.29_{-0.23}^{+0.02}$, and a total hydrogen column density of
$\log N(\mathrm{H})=17.54_{-0.25}^{+0.04}$.
\item[2.] We detect additional narrow components at 8\,\kms, and
58\,\kms\ redward of the \ion{O}{3}-bearing gas (at $v=0$ and
+50\,\kms\ in the $z=0.4925$ rest-frame, respectively). These
components are detected in higher-ionization species. The
$v=0$\,\kms\ component is detected as a single narrow feature in
\ion{N}{4}, \ion{O}{4}, and \ion{S}{4}, while the $v=+50$\,\kms\
component is similarly detected only in \ion{O}{6}. Absorption from
\ion{O}{5}\ and \ion{Ne}{8}\ is detected at the same velocities, but
the kinematics rule out a direct association. The \ion{O}{6}\
profile is too narrow to be produced by collisional processes. Using
limits on the \ion{Ne}{8}/\ion{O}{6}\ and \ion{O}{5}/\ion{O}{6}\
column density ratios, we constrain the ionization parameter of the
$v=+50$\,\kms\ component to the range $-0.35 \lesssim \log U
\lesssim -0.02$.
\item[3.] Two broad, smooth components are also detected in \ion{N}{4},
\ion{O}{5-VI}\ and \ion{Ne}{8}. We find no evidence for the dilution
of troughs expected from gas arising in an accretion-disk outflow.
There is insufficient information to propose a unique ionization
scenario (both collisions and photons could contribute to the
ionization). Radiative cooling and shock ionization models are
unable to produce sufficient amounts of \ion{O}{5} in the ranges of
model parameters considered by \citet{heck02} and \citet{ds96} to
match the observed column density. Photoionization by the quasar is
able to explain some \ion{N}{4} and \ion{O}{4} in a phase tuned to
produce \ion{O}{5-VI}. However, other phases are required to fully
explain the observed profiles. A separate phase is required,
regardless of mechanism, to explain the presence of \ion{Ne}{8}.
\end{itemize}

\acknowledgements The STIS observations of HE0226-4110 were obtained
under the auspices of {\it HST} program 9184, with financial support
through NASA grant HST-GO-9184.08-A from the Space Telescope Science
Institute. TMT was also supported in part by NASA grant NNG 04GG73G.
RG is grateful to Toru Misawa, and Chien Yi Peng for helpful
discussions. The authors also thank the anonymous referee for a
thorough review which resulted in a more streamlined presentation.


\end{document}

%% file: tab1.tex
\begin{deluxetable*}{cccclcc}
\tablewidth{0pc}
\tablecaption{Parameters of Fit to Optical Spectrum}
\tablehead { \colhead{Comp.} & \colhead{$w_{\mathrm{i}}$\tablenotemark{a}} & \colhead{$\lambda_{\mathrm{i}}$} & \colhead{FWHM$_{\mathrm{i}}$\tablenotemark{b}} & \colhead{Ident.} & \colhead{$\zem$} & \colhead{$\log \lambda L_\lambda$\tablenotemark{c}}\\
                             &                                             & \colhead{(\AA)}                  & \colhead{(\kms)}                               &                  &                  & \colhead{(erg s$^{-1}$)}}
\startdata
1 & $0.22 \pm 0.02$ & $7199.6 \pm 2.9$ & $1968 \pm 178$ & UID                        & \nodata             & $43.33 \pm 0.05$ \\
2 & $0.45 \pm 0.07$ & $7259.7 \pm 1.0$ & $ 710 \pm 124$ & H$\beta$                   & $0.4929 \pm 0.0001$ & $43.21 \pm 0.10$ \\
3 & $1.06 \pm 0.06$ & $7263.4 \pm 1.0$ & $2376 \pm 117$ & H$\beta$                   & $0.4937 \pm 0.0001$ & $44.10 \pm 0.03$ \\
4 & $0.25 \pm 0.01$ & $7356.5 \pm 3.1$ & $4610 \pm 414$ & \ion{Fe}{2}                & \nodata             & $43.78 \pm 0.04$ \\
5 & $0.07 \pm 0.02$ & $7358.4 \pm 0.8$ & $ 319 \pm  85$ & UID\tablenotemark{d}       & \nodata             & $42.09 \pm 0.16$ \\
6 & $0.13 \pm 0.02$ & $7474.0 \pm 0.8$ & $ 466 \pm  86$ & [\ion{O}{3}] $\lambda5007$ & $0.4928 \pm 0.0001$ & $42.51 \pm 0.11$ \\
7 & $0.18 \pm 0.01$ & $7481.6 \pm 2.0$ & $2861 \pm 191$ & [\ion{O}{3}] $\lambda5007$ & $0.4942 \pm 0.0002$ & $43.42 \pm 0.04$
\enddata
\tablecomments{Uncertainties are quoted at $1\sigma$\ confidence.
The best-fit flux normalization at 7000\,\AA\ is
$F_{\lambda_{\mathrm{o}}}=(2.125 \pm 0.05) \times 10^{-15}$\, erg
cm$^{-2}$\ s$^{-1}$\ \AA$^{-1}$. The best-fit power-law index was
$\beta=-1.07 \pm 0.09$\ ($F_\lambda \propto \lambda^\beta$). See
\S\ref{sec:quasar} for explanations of the parameters.}
\tablenotetext{a}{Emission line strengths are quoted relative to
best-fit flux normalization at 7000\,\AA.}
\tablenotetext{b}{The width of the emission line components is
quoted as the full-width at half-maximum intensity
($2.35\sigma_{\mathrm{i}}$).}
\tablenotetext{c}{The line luminosity for each component was computed assuming isotropic radiation
and a luminosity distance of 2774.9\,Mpc ($z=0.493$\ for a $\Omega_\Lambda=0.73$, $\Omega_\mathrm{m}=0.27$ cosmology).}
\tablenotetext{d}{One possible identification of this emission line component is [\ion{O}{3}] $\lambda$4959. However,
this is not consistent with either the centroids or strengths of the listed [\ion{O}{3}] $\lambda$5007 components. There
are potential systematic effects resulting from an improper accounting of the underlying \ion{Fe}{2}\ emission, but
such an accounting is beyond the scope of this work.}
\label{tab:emission}
\end{deluxetable*}

%% file: tab2.tex
\begin{deluxetable*}{crrrrrrrr}
\tablewidth{0pc}
\tablecaption{Integration over Narrow Components (in the $\zabs=0.4925$\ rest-frame)}
\tabletypesize{\footnotesize}

\tablehead {
&
&
&
\multicolumn{2}{c}{$-30 \leq v$[\kms]$ \leq +5$} &
\multicolumn{2}{c}{$-30 \leq v$[\kms]$ \leq +30$} &
\multicolumn{2}{c}{$+30 \leq v$[\kms]$ \leq +60$} \\
&
&
&
\multicolumn{2}{c}{\hrulefill} &
\multicolumn{2}{c}{\hrulefill} &
\multicolumn{2}{c}{\hrulefill} \\
\colhead{Ion} &
\colhead{Wavelength\tablenotemark{a}} &
\colhead{$\log f\lambda$\tablenotemark{a}} &
\colhead{$W_\lambda$} &
\colhead{$\acd$\tablenotemark{b,c}} &
\colhead{$W_\lambda$} &
\colhead{$\acd$} &
\colhead{$W_\lambda$} &
\colhead{$\acd$\tablenotemark{d}} \\
&
\colhead{(\AA)}
&
&
\colhead{(m\AA)} &
\colhead{(10$^{13}$\ cm$^{-2}$)} &
\colhead{(m\AA)} &
\colhead{(10$^{13}$\ cm$^{-2}$)} &
\colhead{(m\AA)} &
\colhead{(10$^{13}$\ cm$^{-2}$)}}
\startdata
\ion{H}{1}  & 1025.722 & 1.909 & $149 \pm 4$ &    $\approx 42$ &     \nodata &       \nodata &       $<21$ &         $<1.8$ \\
            &  972.537 & 1.450 & $101 \pm 5$ &         \nodata &     \nodata &       \nodata &     \nodata &        \nodata \\
            &  949.743 & 1.122 & $ 48 \pm 8$ &         \nodata &     \nodata &       \nodata &     \nodata &        \nodata \\ \hline \\[-6pt]
\ion{C}{3}  &  977.020 & 2.869 & $125 \pm 7$ &   $\approx 6.2$ &     \nodata &       \nodata &       $<26$ &        $<0.26$ \\ \hline \\[-6pt]
\ion{N}{4}  &  765.148 & 2.684 &     \nodata &  $\lesssim 2.1$ & $119 \pm 4$ & $4.6 \pm 0.3$ &       $<12$ &        $<0.26$ \\ \hline \\[-6pt]
\ion{O}{1}  &  877.879 & 1.714 &       $<19$ &          $<2.8$ &     \nodata &       \nodata &       $<17$ &         $<2.8$ \\
\ion{O}{2}  &  833.329 & 2.097 &       $<20$ &          $<1.3$ &     \nodata &       \nodata &       $<19$ &         $<1.2$ \\
\ion{O}{3}  &  832.927 & 1.950 & $ 54 \pm 5$ &   $8.4 \pm 1.1$ &     \nodata &       \nodata &       $<18$ &         $<1.9$ \\
\ion{O}{4}  &  787.711 & 1.942 &     \nodata &   $\lesssim 11$ & $142 \pm 5$ &    $36 \pm 3$ &       $<15$ &         $<1.8$ \\
\ion{O}{5}  &  629.730 & 2.511 &     \nodata &         \nodata & $164 \pm 4$ &    $15 \pm 2$ & $66 \pm  4$ & $\lesssim 2.9$ \\
\ion{O}{6}  & 1031.926 & 2.136 &     \nodata &         \nodata & $238 \pm 6$ &    $31 \pm 2$ & $88 \pm  5$ &  $8.7 \pm 1.2$ \\ \hline \\[-6pt]
\ion{Ne}{8} &  770.409 & 1.908 &     \nodata &         \nodata &  $36 \pm 6$ &   $5 \pm 0.8$ & $17 \pm  4$ & $\lesssim 1.8$ \\ \hline \\[-6pt]
\ion{S}{4}  &  748.400 & 2.573 &     \nodata & $\lesssim 0.56$ &  $28 \pm 6$ & $0.9 \pm 0.2$ &       $<14$ &        $<0.35$
\enddata
\tablecomments{Errors on all quantities are quoted at $1\sigma$\ confidence.
Upper limits are quoted at $3\sigma$\ confidence. The velocity range is quoted in {\kms} relative to $\zabs=0.4925$.}
\tablenotetext{a}{Atomic data for transitions above the Lyman limit at 912\,\AA\ were taken from
\citet{morton03}. For transitions blueward of the Lyman limit, we use data compiled by \citet{vbt94}.}
\tablenotetext{b}{The column densities for \ion{H}{1} and \ion{C}{3} were determined using the integrated column
density from the \ion{O}{3} $\lambda$832.927 profile and the column density ratios in Figure~\ref{fig:narrow1a}.}
\tablenotetext{c}{The column density limits for \ion{N}{4}, \ion{O}{4}, and \ion{S}{4} were determined using the
integrated column density from the \ion{O}{3} $\lambda$832.927 profile and the column density limits in
Figure~\ref{fig:narrow1a}.}
\tablenotetext{d}{The column density limits for \ion{O}{5} and \ion{Ne}{8} were determined using the integrated column
density from the \ion{O}{6} 1037.617\,\AA\ profiles and the limiting column density ratios in Figure~\ref{fig:narrow2}.}
\label{tab:narrow}
\end{deluxetable*}

%% file: tab3.tex
\begin{deluxetable*}{crrrrrr}
\tablewidth{0pc} \tablecaption{Integration over Broad Components (in
the $\zabs=0.4925$\ rest-frame)}

\tablehead {
&
&
&
\multicolumn{2}{c}{$-140 \leq v$[\kms]$ \leq -30$} &
\multicolumn{2}{c}{$+60 \leq v$[\kms]$ \leq +200$} \\
&
&
&
\multicolumn{2}{c}{\hrulefill} &
\multicolumn{2}{c}{\hrulefill} \\
\colhead{Ion} &
\colhead{Transition\tablenotemark{a}} &
\colhead{$\log f\lambda$\tablenotemark{a}} &
\colhead{$W_\lambda$} &
\colhead{$\acd$} &
\colhead{$W_\lambda$} &
\colhead{$\acd$} \\
&
\colhead{(\AA)} &
&
\colhead{(m\AA)} &
\colhead{(cm$^{-2}$)} &
\colhead{(m\AA)} &
\colhead{(cm$^{-2}$)} }
\startdata
\ion{H}{1}                  & 1025.722 & 1.909 &      $<44$ &        $<3.5\times10^{13}$ &      $<53$ &        $<4.3\times10^{13}$ \\ \hline \\[-6pt]
\ion{C}{3}                  &  977.020 & 2.869 &      $<49$ &        $<4.5\times10^{12}$ &      $<69$ &        $<5.7\times10^{12}$ \\ \hline \\[-6pt]
\ion{N}{4}                  &  765.148 & 2.684 &  $30\pm 8$ & $(8.0\pm1.7)\times10^{12}$ &  $40\pm11$ & $(9.7\pm1.9)\times10^{12}$ \\ \hline \\[-6pt]
\ion{O}{1}                  &  877.879 & 1.714 &      $<36$ &        $<5.2\times10^{13}$ &      $<42$ &        $<5.9\times10^{13}$ \\
\ion{O}{2}                  &  833.329 & 2.097 &      $<38$ &        $<2.5\times10^{13}$ &      $<46$ &        $<2.7\times10^{13}$ \\
\ion{O}{3}                  &  832.927 & 1.950 &      $<42$ &        $<3.4\times10^{13}$ &      $<45$ &        $<3.9\times10^{13}$ \\
\ion{O}{4}                  &  787.711 & 1.942 &  $40\pm10$ & $(6.0\pm1.2)\times10^{13}$ & $69\pm11$ & $(9.2\pm1.4)\times10^{13}$ \\
\ion{O}{5}                  &  629.730 & 2.511 & $182\pm10$ & $(1.5\pm0.2)\times10^{14}$ & $206\pm12$ & $(1.4\pm0.1)\times10^{14}$ \\
\ion{O}{6}                  & 1031.926 & 2.136 & $201\pm13$ & $(1.8\pm0.1)\times10^{14}$ & $171\pm16$ & $(1.2\pm0.1)\times10^{14}$ \\
\ion{Ne}{8}                 &  770.409 & 1.908 &  $32\pm 8$ & $(4.2\pm1.0)\times10^{13}$ &  $51\pm10$ & $(6.9\pm1.2)\times10^{13}$ \\ \hline \\[-6pt]
\ion{S}{4}                  &  748.400 & 2.573 &      $<26$ &        $<6.7\times10^{12}$ &      $<31$ &        $<7.9\times10^{12}$
\enddata
\tablecomments{Errors on all quantities are quoted at $1\sigma$\ confidence.
Upper limits are quoted at $3\sigma$\ confidence. 
}
\tablenotetext{a}{Atomic data for transitions above the Lyman limit at 912\,\AA\ were taken from
\citet{morton03}. For transitions blueward of the Lyman limit, we use data compiled by \citet{vbt94}.}
%
\label{tab:broad}
\end{deluxetable*}

%% file: tab4.tex
\begin{deluxetable*}{rrr}
\tablewidth{0pc} \tablecaption{Comparison of High-Ionization Species
with Nonequilibrium Models}

\tablehead { & \colhead{$N$(\ion{Ne}{8})/$N$(\ion{O}{6})} &
\colhead{$N$(\ion{O}{5})/$N$(\ion{O}{6})} }

\startdata
$-140 \leq v$[\kms]$ \leq -30$     & $0.23\pm0.06$ & $0.83\pm0.12$ \\
$+60 \leq v$[\kms]$ \leq +200$     & $0.58\pm0.11$ & $1.17\pm0.13$ \\ \hline
Isobaric Cooling\tablenotemark{a}  & 1.26          & 0.5           \\
Isochoric Cooling\tablenotemark{a} & 0.5           & 0.6           \\ \hline
Shock Ionization\tablenotemark{b}  & 0.005--1.026   & 0.097--0.433
\enddata
\tablecomments{Errors on all quantities are quoted at $1\sigma$\
confidence. 
}
\tablenotetext{a}{From \citet{heck02}, these ratios are based on radiative cooling models
from \citet{sd93}, assuming a shock velocity of 600\,\kms, a post-shock velocity of
100\,\kms, and an initial temperature of $10^6$\,K.}
\tablenotetext{b}{From \citet{ds96}, these ratios assume shock velocities in the range
100-500\,\kms, and magnetic parameters in the range 0-4\,$\mu$G cm$^{3/2}$.}
\label{tab:cine}
\end{deluxetable*}